\documentclass[english,aps,prb, twocolumn, showpacs, amsmath, amssymb]{revtex4}
\usepackage[latin1]{inputenc}
\usepackage{prettyref}
\usepackage{amsmath}
\usepackage{graphicx}
\usepackage{amssymb}

\makeatletter

\providecommand{\tabularnewline}{\\}

\usepackage{graphicx}
\usepackage{color}

\usepackage{babel}
\makeatother
\begin{document}

\newcommand{\commute}[2]{\left[#1,#2\right]}

\newcommand{\bra}[1]{\left\langle #1\right|}

\newcommand{\ket}[1]{\left|#1\right\rangle }

\newcommand{\anticommute}[2]{\left\{  #1,#2\right\}  }


\title{Hyperfine interaction in a quantum dot: Non-Markovian electron spin
dynamics}

\author{W. A. Coish}

\author{Daniel Loss}

\affiliation{Department of Physics and Astronomy, University of Basel, Klingelbergstrasse
82, CH-4056 Basel, Switzerland }

\begin{abstract}
We have performed a systematic calculation for the non-Markovian dynamics
of a localized electron spin interacting with an environment of nuclear
spins via the Fermi contact hyperfine interaction. This work applies
to an electron in the $s$-type orbital ground state of a quantum
dot or bound to a donor impurity, and is valid for arbitrary polarization
$p$ of the nuclear spin system, and arbitrary nuclear spin $I$ in
high magnetic fields. In the limit of $p=1$ and $I=\frac{1}{2}$,
the Born approximation of our perturbative theory recovers the exact
electron spin dynamics. We have found the form of the generalized
master equation (GME) for the longitudinal and transverse components
of the electron spin to all orders in the electron spin--nuclear spin
flip-flop terms. Our perturbative expansion is regular, unlike standard
time-dependent perturbation theory, and can be carried-out to higher
orders. We show this explicitly with a fourth-order calculation of
the longitudinal spin dynamics. In zero magnetic field, the fraction
of the electron spin that decays is bounded by the smallness parameter
$\delta=1/p^{2}N$, where $N$ is the number of nuclear spins within
the extent of the electron wave function. However, the form of the
decay can only be determined in a high magnetic field, much larger
than the maximum Overhauser field. In general the electron spin shows
rich dynamics, described by a sum of contributions with non-exponential
decay, exponential decay, and undamped oscillations. There is an abrupt
crossover in the electron spin asymptotics at a critical dimensionality
and shape of the electron envelope wave function. We propose a scheme
that could be used to measure the non-Markovian dynamics using a standard
spin-echo technique, even when the fraction that undergoes non-Markovian
dynamics is small. 
\end{abstract}
\maketitle

\section{Introduction}

Prospects for the development of new spintronic devices,\cite{wolf:2001a}
and the controlled manipulation of electron or nuclear spins for quantum
information processing\cite{awschalom:2002a} have sparked substantial
research efforts in recent years. One of the major obstacles to achieving
these goals is decoherence due to the influence of an uncontrollable
environment. For quantum computing tasks, the strict requirements
for error correction\cite{preskill:1998a} put strong limits on the
degree of decoherence allowed in such devices. From this point of
view, single-electron semiconductor quantum dots represent good candidates
for spin-based information processing since they show particularly
long longitudinal relaxation times, $T_{1}=1\,\mathrm{ms}$.\cite{kouwenhoven:2004a}
In GaAs quantum wells, the transverse dephasing time $T_{2}^{*}$
for an ensemble of electron spins, which typically provides a lower
bound for the intrinsic decoherence time $T_{2}$ of an isolated spin,
has been measured to be in excess of $100\,\mathrm{ns}$.\cite{kikkawa:1998a} 

Possible sources of decoherence for a single electron spin confined
to a quantum dot are spin-orbit coupling and the contact hyperfine
interaction with the surrounding nuclear spins.\cite{burkard:1999a}
The relaxation rate due to spin-orbit coupling $\frac{1}{T_{1}}$
is suppressed for localized electrons at low temperatures\cite{khaetskii:2000a,khaetskii:2001a}
and recent work has shown that $T_{2}$, due to spin-orbit coupling,
can be as long as $T_{1}$ under realistic conditions.\cite{golovach:2003a}
However, since spin-carrying isotopes are common in the semiconductor
industry, the contact hyperfine interaction (in contrast to the spin-orbit
interaction) is likely an unavoidable source of decoherence, which
does not vanish with decreasing temperature or carefully chosen quantum
dot geometry.\cite{schliemann:2003a} 

In the last few years, a great deal of effort has been focused on
a theoretical description of interesting effects arising from the
contact hyperfine interaction for a localized electron.\cite{burkard:1999a,khaetskii:2002a,khaetskii:2003a,schliemann:2002a,lyanda-geller:2002a,taylor:2003a,taylor:2003b,imamoglu:2003a,semenov:2004a,erlingsson:2001a,erlingsson:2002a,desousa:2003a,desousa:2003b,semenov:2003a,merkulov:2002a,saykin:2002a}
The predicted effects include a dramatic variation of $T_{1}$ with
gate voltage in a quantum dot near the Coulomb blockade peaks or valleys,\cite{lyanda-geller:2002a}
all-optical polarization of the nuclear spins,\cite{imamoglu:2003a}
use of the nuclear spin system as a quantum memory,\cite{taylor:2003a,taylor:2003b}
and several potential spin relaxation and decoherence mechanisms.\cite{khaetskii:2002a,erlingsson:2001a,erlingsson:2002a,semenov:2004a,desousa:2003a}
This theoretical work is spurred-on by intriguing experiments that
show localized electrical detection of spin resonance phenomena,\cite{dobers:1988a}
nuclear spin polarization near quantum point contacts,\cite{wald:1994a}
gate-controlled transfer of polarization between electrons and nuclei,\cite{smet:2002a}
nuclear spin polarization and manipulation due to optical pumping
in GaAs quantum wells,\cite{salis:2001a} and voltage-controlled nuclear
spin polarization in a field-effect transistor.\cite{epstein:2003a}
In addition, recent experiments have shown hyperfine induced oscillations
in transport current through a double quantum dot,\cite{ono:2003a}
and long $T_{2}$ times for electrons trapped at shallow donor impurities
in isotopically purified $^{28}$Si:P.\cite{tyryshkin:2003a} Our
system of interest in this paper is an electron confined to a single
GaAs quantum dot, but this work applies quite generally to other systems,
such as electrons trapped at shallow donor impurities in Si:P.\cite{schliemann:2003a}

In this paper, we investigate electron spin dynamics at times shorter
than the nuclear dipole-dipole correlation time $\tau_{\mathrm{dd}}$
($\tau_{\mathrm{dd}}\approx10^{-4}\,\mathrm{s}$ in GaAs is given
directly by the inverse width of the nuclear magnetic resonance (NMR)
line\cite{paget:1977a}). At these time scales, the relevant Hamiltonian
for a description of the electron and nuclear spin dynamics is that
for the Fermi contact hyperfine interaction (see Eq. \prettyref{eq:HFHamiltonian},
below). Dynamics under the action of this Hamiltonian may be of fundamental
interest, since in zero magnetic field, Eq. \prettyref{eq:HFHamiltonian}
corresponds to the well-known integrable Gaudin magnet, which is soluble
via Bethe ansatz.\cite{gaudin:1976a,schliemann:2003a} Though the
Hamiltonian appears simple, a detailed microscopic description for
the dynamics of a spin coupled to a spin environment remains an open
question.\cite{dobrovitski:2003a,zurek:2003a} A degree of success
has been achieved some time ago in bulk systems through the development
of phenomenological models.\cite{klauder:1962a} These models invoke
certain approximations, namely, assumptions of Markovian dynamics
and ensemble averaging. Care should therefore be taken in applying
the same models to the problem of single-spin decoherence for an electron
spin strongly coupled to a nuclear spin environment, where they may
not apply.\cite{khaetskii:2002a,khaetskii:2003a} 

For nuclear spin $I=\frac{1}{2}$, an exact solution for the electron
spin dynamics has been found in the special case of a \emph{fully}
polarized initial state of the nuclear spin system.\cite{khaetskii:2002a,khaetskii:2003a}
This solution shows that the electron spin only decays by a fraction
$\propto\frac{1}{N}$ of its initial value, where $N$ is the number
of nuclear spins within the extent of the electron wave function.
The decaying fraction was shown to have a non-exponential tail for
long times, which suggests non-Markovian (history dependent) behavior.
For an initial nuclear spin configuration that is not fully polarized,
no exact solution is available and standard time-dependent perturbation
theory fails.\cite{khaetskii:2002a} Subsequent exact diagonalization
studies on small spin systems\cite{schliemann:2002a} have shown that
the electron spin dynamics are highly dependent on the type of initial
nuclear spin configuration, and the dynamics of a randomly correlated
initial nuclear spin configuration are reproduced by an ensemble average
over direct-product initial states. The unusual (non-exponential)
form of decay, and the fraction of the electron spin that undergoes
decay may be of interest in quantum error correction (QEC) since QEC
schemes typically assume exponential decay to zero. 

In this paper we formulate a systematic perturbative theory of electron
spin dynamics under the action of the Fermi contact hyperfine interaction.
This theory is valid for \emph{arbitrary} nuclear spin polarization
and \emph{arbitrary} nuclear spin $I$ in high magnetic fields. For
nuclear spin $I=\frac{1}{2}$ and a fully polarized nuclear spin system,
we recover the exact solution for the electron spin dynamics within
the Born approximation of our perturbative theory. Our approach follows
a method recently applied to the spin-boson model.\cite{loss:2003a}
This method does not suffer from unbounded secular terms that occur
in standard perturbation theory\cite{khaetskii:2002a} and does not
involve Markovian approximations. 

This paper is organized as follows. In \prettyref{sec:model} we review
the model Hamiltonian and address the question of realistic initial
conditions. In \prettyref{sec:Master-equation} we derive the form
of the exact generalized master equation (GME) for the electron spin
dynamics. In \prettyref{sec:High-field-solution} we consider the
leading-order electron spin dynamics in high magnetic fields. In \prettyref{sec:Non-Markovian-dynamics}
we proceed to calculate the complete non-Markovian dynamics within
the Born approximation. We describe a procedure that could be used
to measure the non-Markovian dynamics in \prettyref{sec:Measurement}.
In \prettyref{sec:Beyond-Born} we show that our method can be extended
to higher orders without the problems of standard perturbation theory
by explicitly calculating the corrections to the longitudinal spin
self-energy at fourth order in the nuclear spin--electron spin flip-flop
terms. We conclude in \prettyref{sec:Conclusions} with a summary
of the results. Technical details are deferred to Appendices \ref{sec:Self-energy-expansion}--\ref{sec:Branch-cut-asymptotics}.

\section{Model\label{sec:model}}

\subsection{Hamiltonian}

We consider a localized electron spin interacting with $N_{\mathrm{tot}}$
nuclear spins via the Fermi contact hyperfine interaction. The Hamiltonian
for this system is\begin{equation}
\mathcal{H}=bS_{z}+\epsilon_{nz}I_{z}+\mathbf{h}\cdot\mathbf{S},\label{eq:HFHamiltonian}\end{equation}
where $\mathbf{S}=\left(S_{x},S_{y},S_{z}\right)$ is the electron
spin operator. $b=g^{*}\mu_{B}B_{z}\,(\epsilon_{nz}=g_{I}\mu_{N}B_{z})$
is the electron (nuclear) Zeeman splitting in a magnetic field $B_{z}$,
with effective g-factor $g^{*}$ ($g_{I}$) for the electron (nuclei)
and Bohr (nuclear) magneton $\mu_{B}$ ($\mu_{N}$). Further, $\mathbf{h}=\left(h_{x},h_{y},h_{z}\right)=\sum_{k=0}^{N_{\mathrm{tot}}-1}A_{k}\mathbf{I}_{k}$
gives the (quantum) field generated by an environment of nuclear spins.
$\mathbf{I}_{k}=\left(I_{k}^{x},I_{k}^{y},I_{k}^{z}\right)$ is the
nuclear spin operator at lattice site $k$ and $A_{k}$ is the associated
hyperfine coupling constant. $I_{z}=\sum_{k}I_{k}^{z}$ is the total
$z$-component of nuclear spin.

The nuclear Zeeman term can be formally eliminated from the Hamiltonian
$\mathcal{H}$ (Eq. \prettyref{eq:HFHamiltonian}) by transforming
to a rotating reference frame. The $z$-component of total angular
momentum is $J_{z}=S_{z}+I_{z}$. Adding and subtracting $\epsilon_{nz}J_{z}$
gives $\mathcal{H}=\mathcal{H}^{\prime}+\epsilon_{nz}J_{z}$. The
Hamiltonian in the rotating frame, $\mathcal{H}^{\prime}$, is then\begin{eqnarray}
\mathcal{H}^{\prime} & = & \mathcal{H}_{0}^{\prime}+\mathcal{H}_{V}^{\prime},\label{eq:HinRotatingFrame}\\
\mathcal{H}_{0}^{\prime} & = & b^{\prime}S_{z}+h_{z}S_{z},\label{eq:HamiltonianH0}\\
\mathcal{H}_{V}^{\prime} & = & \frac{1}{2}\left(h_{+}S_{-}+h_{-}S_{+}\right),\label{eq:HamiltonianHV}\end{eqnarray}
 where $b^{\prime}=b-\epsilon_{nz}$ and we have introduced $h_{\pm}=h_{x}\pm ih_{y}$.
The usual Heisenberg-picture operators in the rotating frame are $S_{X}^{\prime}(t)=e^{i\mathcal{H}^{\prime}t}S_{X}e^{-i\mathcal{H}^{\prime}t},\, X=z,+,\, S_{\pm}=S_{x}\pm iS_{y}$.
Noting that $\commute{J_{z}}{\mathcal{H}}=0$, we find they are related
to the operators $S_{X}(t)=e^{i\mathcal{H}t}S_{X}e^{-i\mathcal{H}t}$
in the rest frame by\begin{eqnarray}
S_{z}^{\prime}(t) & = & S_{z}(t)\\
S_{+}^{\prime}(t) & = & e^{-i\epsilon_{nz}t}S_{+}(t).\end{eqnarray}
 In the following, $\left\langle S_{z}^{\prime}\right\rangle _{t}$
and $\left\langle S_{+}^{\prime}\right\rangle _{t}$ will be evaluated
in the rotating frame, but we omit primes on all expectation values.

\begin{figure}
\includegraphics[%
  bb=60bp 570bp 550bp 780bp,
  clip,
  scale=0.5]{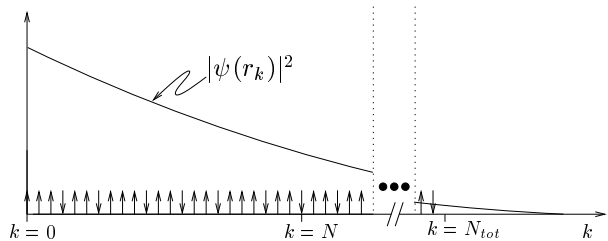}

\caption{\label{cap:Wavefunction}Schematic of the square modulus of the electron
envelope wave function $\left|\psi(r)\right|^{2}$ and nuclear spins
(arrows). $k$ is the nuclear site index, $N$ is the number of nuclear
spins within radius $r=l_{0}$, and $N_{\mathrm{tot}}$ is the total
number of nuclear spins in the system.}
\end{figure}

The hyperfine coupling constants $A_{k}$ are given by\cite{schliemann:2003a}\begin{equation}
A_{k}=Av_{0}|\psi(\mathbf{r}_{k})|^{2}.\end{equation}
 Here, $v_{0}$ is the volume of a crystal unit cell containing one
nuclear spin, $\psi(\mathbf{r})$ is the electron envelope wave function,
and $A$ is the strength of the hyperfine coupling. In GaAs, all naturally
occurring isotopes carry spin $I=\frac{3}{2}$. In bulk GaAs, $A$
has been estimated\cite{paget:1977a} to be $A=90\,\mu eV$ $\left(\frac{A}{|g^{*}|\mu_{B}}=3.5\,\mathrm{T}\right)$.
This estimate is based on an average over the hyperfine coupling constants
for the three nuclear isotopes $\mathrm{^{69}Ga}$, $\mathrm{^{71}Ga}$,
and $\mathrm{^{75}As}$, weighted by their relative abundances. Natural
silicon contains 4.7\% $^{29}\mathrm{Si}$, which carries $I=\frac{1}{2}$,
and 95\% $^{28}\mathrm{Si}$, with $I=0$. An electron bound to a
phosphorus donor impurity in natural Si:P interacts with $N\approx10^{2}$
surrounding $\mathrm{^{29}Si}$ nuclear spins, in which case the hyperfine
coupling constant is on the order of $A\approx0.1\,\mu eV$.\cite{schliemann:2003a}
We consider a localized electron in its orbital ground state, described
by an isotropic envelope wave function of the form \begin{equation}
\psi(r_{k})=\psi(0)\exp\left[-\frac{1}{2}\left(\frac{r_{k}}{l_{0}}\right)^{m}\right].\end{equation}
 When $m=2$, $\psi(r)$ is a Gaussian with Bohr radius $l_{0}$,
and for $m=1$, $\psi(r)$ corresponds to a hydrogen-like $s$-state
with Bohr radius $a_{0}=2l_{0}$. $N_{\mathrm{tot}}$ nuclear spins
are in the system, but the effective number $N$ of spins interacting
appreciably with the electron is smaller (see Fig. \ref{cap:Wavefunction}).
$N$ is defined as the number of nuclear spins within radius $l_{0}$
of the origin and the integer index $k$ gives the number of spins
within radius $r_{k}$. In $d$ dimensions, $\left(\frac{r_{k}}{l_{0}}\right)^{d}=\frac{k}{N}$.
It is convenient to work in energy units such that $\frac{A_{0}}{2}=1$,
where $A_{0}$ is the coupling constant at the origin $(r_{0}=0)$.
In these units $A_{k}$ takes the simple form\begin{equation}
A_{k}=2\exp\left[-\left(\frac{k}{N}\right)^{\frac{m}{d}}\right].\label{eq:AkDefinition}\end{equation}

\subsection{Initial conditions}

\subsubsection{Sudden approximation}

The electron spin and nuclear system are decoupled for times $t<0$,
and prepared independently in states described by the density operators
$\rho_{S}(0)$ and $\rho_{I}(0)$, respectively. At $t=0$, the electron
and nuclear spin system are brought into contact {}``instantaneously'',
i.e., the electron spin and nuclear system are brought into contact
over a switching time scale $\tau_{\mathrm{sw}}$,%
\footnote{$\tau_{\mathrm{sw}}$ is, e.g., the time taken to inject an electron
into a quantum dot.%
} which is sufficiently small--see Eq. \prettyref{eq:SwitchingTime},
below. The state of the entire system, described by the total density
operator $\rho(t)$ is then continuous at $t=0$, and is given by\begin{equation}
\rho(0^{-})=\rho(0^{+})=\rho_{S}(0)\otimes\rho_{I}(0).\label{eq:SuddenApproximationProductState}\end{equation}
 The evolution of the density operator $\rho(t)$ for $t\ge0$ is
governed by the Hamiltonian $\mathcal{H}^{\prime}$ for an electron
spin coupled to an environment of nuclear spins. Since the largest
energy scale in this problem is given by $\left|b^{\prime}+A\right|$,
in general the condition\begin{equation}
\tau_{\mathrm{sw}}\ll\frac{2\pi\hbar}{\left|b^{\prime}+A\right|}\label{eq:SwitchingTime}\end{equation}
should be satisfied for the sudden approximation (Eq. \prettyref{eq:SuddenApproximationProductState})
to be valid. In bulk GaAs, $\frac{2\pi\hbar}{A}\simeq50\,\mathrm{ps}$
and for an electron bound to a phosphorus donor in natural silicon,
$\frac{2\pi\hbar}{A}\simeq10\,\mathrm{ns}$.

\subsubsection{Dependence on the nuclear state: zeroth order dynamics}

Evolution of the electron spin for different initial nuclear configurations
has been addressed previously.\cite{khaetskii:2003a,schliemann:2002a}
In Ref. \onlinecite{schliemann:2002a} it was found, through numerical
study, that the dynamics of the electron spin were highly dependent
on the initial state of the nuclear system. The goal of this section
is to shed more light on the role of the initial nuclear configuration
by evaluating the much simpler zeroth order dynamics, i.e., the electron
spin evolution is evaluated under $\mathcal{H}^{\prime}=\mathcal{H}_{0}^{\prime}$
alone, neglecting the flip-flop terms $\mathcal{H}_{V}^{\prime}$. 

Since $\commute{\mathcal{H}_{0}^{\prime}}{S_{z}}=0$, $\left\langle S_{z}\right\rangle _{t}$
is constant. However, $\commute{\mathcal{H}_{0}^{\prime}}{S_{\pm}}\ne0$,
so the transverse components, $\left\langle S_{+}\right\rangle _{t}=\left\langle S_{x}\right\rangle _{t}+i\left\langle S_{y}\right\rangle _{t}$,
will have a nontrivial time dependence. We evaluate the expectation
value $\left\langle S_{+}\right\rangle _{t}=\mathrm{Tr}\left\{ e^{-i\mathcal{H}_{0}^{\prime}t}S_{+}e^{i\mathcal{H}_{0}^{\prime}t}\rho(0)\right\} $
(setting $\hbar=1$), with the initial state given in Eq. \prettyref{eq:SuddenApproximationProductState}.
After performing a partial trace over the electron spin Hilbert space,
we obtain an expression in terms of the initial nuclear spin state:\begin{equation}
\left\langle S_{+}\right\rangle _{t}=\left\langle S_{+}\right\rangle _{0}\mathrm{Tr}_{I}\left\{ e^{i(b^{\prime}+h_{z})t}\rho_{I}(0)\right\} ,\label{eq:SplusTimeEvolution}\end{equation}
 where $\mathrm{Tr}_{I}$ is a partial trace over the nuclear spin
space alone. For simplicity, here we consider $I=\frac{1}{2}$, and
the coupling constants are taken to be uniform. After enforcing the
normalization $\sum_{k}A_{k}=2N$ in units where $\frac{A_{0}}{2}=\frac{A}{2N}=1$,
the hyperfine coupling constants are

\begin{equation}
A_{k}=\left\{ \begin{array}{l}
2,\,\,\, k=0,1\cdots N-1\\
0,\,\,\, k\ge N\end{array}\right..\end{equation}
The zeroth-order electron spin dynamics can now be evaluated exactly
for three types of initial nuclear spin configuration:

\begin{eqnarray}
\rho_{I}^{(1)}(0) & = & \ket{\psi_{I}(0)}\bra{\psi_{I}(0)}\\
\rho_{I}^{(2)}(0) & = & \sum_{N_{\uparrow}=0}^{N}P(N_{\uparrow};N,f_{\uparrow})\ket{N_{\uparrow}}\bra{N_{\uparrow}}\\
\rho_{I}^{(3)}(0) & = & \ket{n}\bra{n}.\label{eq:InitialNuclearStates}\end{eqnarray}
 $\rho_{I}^{(1)}$ is a pure state, where $\ket{\psi_{I}(0)}=\prod_{k=0}^{N}\left(\sqrt{f_{\uparrow}}\ket{\uparrow_{k}}+e^{i\phi_{k}}\sqrt{1-f_{\uparrow}}\ket{\downarrow_{k}}\right)$
is chosen to render the z-component of nuclear spin translationally
invariant: $\bra{\psi_{I}(0)}I_{k}^{z}\ket{\psi_{I}(0)}=\frac{1}{2}(2f_{\uparrow}-1)=\frac{p}{2}$
, and $p=2f_{\uparrow}-1$ is the polarization of the nuclear spin
system. $\phi_{k}$ is an arbitrary site-dependent phase factor. $P(x;n,f)=\left(\begin{array}{c}
n\\
x\end{array}\right)f^{x}(1-f)^{n-x}$ is a binomial distribution, and $\ket{N_{\uparrow}}$ is a product
state of the form $\ket{\uparrow\uparrow\downarrow\cdots}$ with $N_{\uparrow}$
spins up and $N-N_{\uparrow}$ spins down. $\rho_{I}^{(2)}(0)$ then
corresponds to a mixed state; this is an ensemble of product states
where the $N$ spins in each product state are selected from a bath
of polarization $p=2f_{\uparrow}-1$. $\rho_{I}^{(3)}$, like $\rho_{I}^{(1)}$,
is a pure state, but for this state $\ket{n}$ is chosen to be an
eigenstate of $h_{z}$ with eigenvalue $pN$ (corresponding to a nuclear
system with polarization $p$): $h_{z}\ket{n}=pN\ket{n}$. We insert
the initial nuclear spin states $\rho_{I}^{(i)}(0)$ into \prettyref{eq:SplusTimeEvolution}
to obtain the associated time evolution $\left\langle S_{+}\right\rangle _{t}^{(i)}$:\begin{eqnarray}
\left\langle S_{+}\right\rangle _{t}^{(1,2)} & =\left\langle S_{+}\right\rangle _{0} & \sum_{N_{\uparrow}=0}^{N}P(N_{\uparrow};N,f_{\uparrow})e^{i(b^{\prime}+M(N_{\uparrow}))t},\,\,\\
\left\langle S_{+}\right\rangle _{t}^{(3)} & = & \left\langle S_{+}\right\rangle _{0}\mathrm{e}^{i(b^{\prime}+pN)t}.\end{eqnarray}
 $M(N_{\uparrow})=2N_{\uparrow}-N$ is the nuclear magnetization on
a dot with $N_{\uparrow}$ nuclear spins up.

The similarity in dynamics between randomly correlated (entangled)
pure states and mixed states has been demonstrated for evolution under
the full Hamiltonian $(\mathcal{H}^{\prime}=\mathcal{H}_{0}^{\prime}+\mathcal{H}_{V}^{\prime})$
via exact diagonalizations of small ($N_{\mathrm{tot}}\lesssim19$)
spin systems.\cite{schliemann:2002a} Here, the zeroth order electron
spin dynamics are identical for the pure state $\rho_{I}^{(1)}(0)$
and the mixed state $\rho_{I}^{(2)}(0)$ even when the initial pure
state $\ket{\psi_{I}(0)}$ is a direct product. Direct application
of the central limit theorem gives a Gaussian decay for large $N$:\begin{equation}
\left\langle S_{+}\right\rangle _{t}^{(1,2)}\approx\left\langle S_{+}\right\rangle _{0}e^{-\frac{t^{2}}{2t_{\mathrm{c}}^{2}}+i(b^{\prime}+pN)t},\, t_{\mathrm{c}}=\frac{1}{\sqrt{N(1-p^{2})}}.\label{eq:GaussianDecay}\end{equation}
 Returning to dimension-full units (c.f. Table \prettyref{cap:Symbols}
below), the time scale for this decay is given by $\tau_{\mathrm{c}}=\frac{2N\hbar}{A}t_{\mathrm{c}}\approx5\,\mathrm{ns}$
for a GaAs quantum dot with $p^{2}\ll1$ containing $N=10^{5}$ nuclei
and $\tau_{c}\approx100\,\mathrm{ns}$ for an electron trapped at
a shallow donor impurity in $\mathrm{Si:P}$, with $N=10^{2}$. For
an ensemble of nuclear spin states, Gaussian decay with the time scale
$\tau_{c}$ has been found previously.\cite{khaetskii:2002a,khaetskii:2003a,merkulov:2002a}
Gaussian decay for a Hamiltonian with an Ising coupling of electron
and nuclear spins has been demonstrated\cite{zurek:2003a} for a more
general class of $pure$ initial states and for coupling constants
$A_{k}$ that may vary from site-to-site.

For the initial states $\rho_{I}^{(1,2)}(0)$, precise control over
the nuclear spin polarization between measurements or a spin-echo
technique would be needed to reduce or eliminate the rapid decay described
by \prettyref{eq:GaussianDecay}. However, the quantum superposition
of $h_{z}$ eigenstates can be removed, in principle, from the pure
state $\rho_{I}^{(1)}(0)$ by performing a strong (von Neumann) measurement
on the nuclear Overhauser field $pN$.%
\footnote{It may be possible to measure the Overhauser field directly by locating
the position of the electron spin resonance (ESR) line, where the
magnetic field compensates the nuclear Overhauser field. We have confirmed
by exact diagonalizations on small ($N_{\mathrm{tot}}=15$) spin systems
that the resonance is indeed centered at a magnetic field corresponding
to the negative nuclear Overhauser field, even for a nuclear spin
system with $p<1$. Alternatively, a state where all nuclear spins
are aligned along the magnetic field can be generated by allowing
the nuclear spins to relax in the presence of the nuclear spin-lattice
interaction.%
} After the nuclear system is prepared in an $h_{z}$-eigenstate, to
zeroth order the electron spin dynamics will be given by $\left\langle S_{+}\right\rangle _{t}^{(3)}$,
i.e., a simple precession about the z-axis with no decay. 

When higher-order corrections are taken into account, and the coupling
constants $A_{k}$ are allowed to vary from site-to-site, even an
initial $h_{z}$-eigenstate can lead to decay of the electron spin.
This has been shown\cite{khaetskii:2002a,khaetskii:2003a} in an exact
solution for the specific case of a fully-polarized system of nuclear
spins-$\frac{1}{2}$ and by exact diagonalization on small systems.\cite{schliemann:2002a}
The goal of the present work is to perform an analytical calculation
with a larger range of validity (a large system of nuclear spins with
arbitrary polarization and arbitrary nuclear spin $I$ in a sufficiently
strong magnetic field) that recovers previous exact results in the
relevant limiting cases. In the rest of this paper, the effect of
higher- (beyond zeroth-) order corrections will be considered for
a nuclear spin system prepared in an arbitrary $h_{z}$ eigenstate:
$\rho_{I}(0)=\rho_{I}^{(3)}(0)$, as given in Eq. \prettyref{eq:InitialNuclearStates}.
Specifically, the initial state of the nuclear system $\ket{n}$ can
be written as an arbitrary linear combination of $g_{n}$ degenerate
product states:\begin{equation}
\ket{n}=\sum_{j=1}^{g_{n}}\alpha_{j}\ket{n_{j}},\,\,\,\,\,\ket{n_{j}}=\bigotimes_{i=0}^{N_{\mathrm{tot}}-1}\ket{I,m_{i}^{j}}\label{eq:SpecificInitialState}\end{equation}
 where $\ket{I,m_{i}}$ is an eigenstate of the operator $I_{i}^{z}$
with eigenvalue $m_{i}$ and $h_{z}\ket{n_{j}}=\left[h_{z}\right]_{nn}\ket{n_{j}}$
for all $j$, where we write the matrix elements of any operator $\mathcal{O}$
as $\bra{i}\mathcal{O}\ket{j}=\left[\mathcal{O}\right]_{ij}$.

\section{Generalized master equation\label{sec:Master-equation}}

To evaluate the dynamics of the reduced (electron spin) density operator,
we introduce a projection superoperator $P$, defined by its action
on an arbitrary operator $\mathcal{O}$: $P\mathcal{O}=\rho_{I}(0)\mathrm{Tr}_{I}\mathcal{O}$.
$P$ is chosen to preserve all electron spin expectation values: $\left\langle S_{\beta}\right\rangle _{t}=\mathrm{Tr}S_{\beta}\rho(t)=\mathrm{Tr}S_{\beta}P\rho(t),\,\beta=x,y,z$,
and satisfies $P^{2}=P$. For factorized initial conditions (Eq. \prettyref{eq:SuddenApproximationProductState}),
$P\rho(0)=\rho(0)$, which is a sufficient condition to rewrite the
von Neumann equation $\dot{\rho}(t)=-i\commute{\mathcal{H}^{\prime}}{\rho(t)}$
in the form of the exact Nakajima-Zwanzig generalized master equation
(GME)\cite{fick:1990a}:\begin{eqnarray}
P\dot{\rho}(t) & = & -iPLP\rho(t)-i\int_{0}^{t}dt^{\prime}\Sigma(t-t^{\prime})\rho(t^{\prime}),\label{eq:NZGME}\\
\Sigma(t) & = & -iPLQe^{-iLQt}QLP,\end{eqnarray}
 where $\Sigma(t)$ is the self-energy superoperator and $Q=\mathbf{1}-P$
is the complement of $P$ ($\mathbf{1}$ is the identity superoperator).
$L=L_{0}+L_{V}$ is the full Liouvillian, where $L_{\alpha}$ $(\alpha=V,0)$
is defined by $L_{\alpha}\mathcal{O}=\commute{\mathcal{H}_{\alpha}^{\prime}}{\mathcal{O}}$.
When the initial nuclear state is of the form $\rho_{I}(0)=\ket{n}\bra{n}$,
where $\ket{n}$ is an arbitrary eigenstate of $h_{z}$, as in Eq.
\prettyref{eq:SpecificInitialState}, $P$ obeys the useful identities\begin{eqnarray}
PL_{V}P & = & 0,\label{eq:PLvPZero}\\
PL_{0}P & = & L_{0}P.\label{eq:PL0PZero}\end{eqnarray}
We apply Eqs. \prettyref{eq:PLvPZero} and \prettyref{eq:PL0PZero},
and perform a trace on \prettyref{eq:NZGME} over the nuclear spins
to obtain\begin{eqnarray}
\dot{\rho}_{S}(t) & = & -iL_{0}^{n}\rho_{S}(t)-i\int_{0}^{t}dt^{\prime}\Sigma_{S}(t-t')\rho_{S}(t'),\\
\Sigma_{S}(t) & = & -i\mathrm{Tr}_{I}Le^{-iQLt}L_{V}\rho_{I}(0),\label{eq:ReducedNZGME}\end{eqnarray}
 where $L_{0}^{n}\mathcal{O}=\commute{S_{z}\omega_{n}}{\mathcal{O}}$
and $\omega_{n}=b^{\prime}+\left[h_{z}\right]_{nn}$. $\Sigma_{S}(t)$
is the reduced self-energy superoperator. $\rho_{S}(t)=\mathrm{Tr}_{I}\rho(t)=\frac{1}{2}\sigma_{0}+\left\langle S_{x}\right\rangle _{t}\sigma_{x}+\left\langle S_{y}\right\rangle _{t}\sigma_{y}+\left\langle S_{y}\right\rangle _{t}\sigma_{y}$
is the reduced electron spin density operator, where $\sigma_{\beta},\,\beta=x,y,z,$
are the usual Pauli matrices and $\sigma_{0}$ is the $2\times2$
identity.

We iterate the Schwinger-Dyson identity \cite{fick:1990a}\begin{equation}
e^{-iQ(L_{0}+L_{V})t}=e^{-iQL_{0}t}-i\int_{0}^{t}dt'e^{-iQL_{0}(t-t')}QL_{V}e^{-iQLt'}\end{equation}
 on \prettyref{eq:ReducedNZGME} to generate a systematic expansion
of the reduced self-energy in terms of the perturbation Liouvillian
$L_{V}$:\begin{equation}
\Sigma_{S}(t)=\Sigma_{S}^{(2)}(t)+\Sigma_{S}^{(4)}(t)+\cdots,\label{eq:Self-EnergyExpansion}\end{equation}
 where the superscript indicates the number of occurrences of $L_{V}$.
Quite remarkably, to all orders in $L_{V}$, the equations for the
longitudinal $\left(\left\langle S_{z}\right\rangle _{t}\right)$
and transverse $\left(\left\langle S_{+}\right\rangle _{t}=\left\langle S_{x}\right\rangle _{t}+i\left\langle S_{y}\right\rangle _{t}\right)$
electron spin components are decoupled and take the form:\begin{eqnarray}
\dot{\left\langle S_{z}\right\rangle }_{t} & = & N_{z}(t)-i\int_{0}^{t}dt^{\prime}\Sigma_{zz}(t-t^{\prime})\left\langle S_{z}\right\rangle _{t^{\prime}}\label{eq:SzGME}\\
\dot{\left\langle S_{+}\right\rangle }_{t} & = & i\omega_{n}\left\langle S_{+}\right\rangle _{t}-i\int_{0}^{t}dt^{\prime}\Sigma_{++}(t-t^{\prime})\left\langle S_{+}\right\rangle _{t^{\prime}}.\label{eq:SPlusGME}\end{eqnarray}
Details of the expansion (Eq. \prettyref{eq:Self-EnergyExpansion})
are given in Appendix \ref{sec:Self-energy-expansion}. It is most
convenient to evaluate the inhomogeneous term $N_{z}(t)$ and the
memory kernels $\Sigma_{zz}(t),\,\Sigma_{++}(t)$ in terms of their
Laplace transforms: $f(s)=\int_{0}^{\infty}dte^{-st}f(t),\,\mathrm{Re}[s]>0$.
$N_{z}(s)$ and $\Sigma_{zz}(s)$ are given in terms of matrix elements
of the reduced self-energy by\begin{eqnarray}
N_{z}(s) & = & -\frac{i}{2s}\left(\Sigma_{\uparrow\uparrow}(s)+\Sigma_{\uparrow\downarrow}(s)\right),\label{eq:NzBA}\\
\Sigma_{zz}(s) & = & \Sigma_{\uparrow\uparrow}(s)-\Sigma_{\uparrow\downarrow}(s).\label{eq:SigzzBA}\end{eqnarray}
Explicit expressions for the matrix elements $\Sigma_{++}(s),\,\Sigma_{\uparrow\uparrow}(s),$
and $\Sigma_{\uparrow\downarrow}(s)$ are given in Appendix \ref{sec:Self-energy-expansion}.
We find that the self-energy at $(2k)^{\mathrm{th}}$ order is suppressed
by the factor $\Delta^{k}$, where\begin{equation}
\Delta=\frac{N}{\omega_{n}}.\end{equation}
 The parameter $\Delta$ and some other commonly used symbols are
given in dimensionless and dimension-full units in Table \prettyref{cap:Symbols}
below. For high magnetic fields $\left|b^{\prime}\right|\gg N$ $\left(\left|B_{z}\right|\gg\left|\frac{A}{g^{*}\mu_{B}}\right|\right)$,
we have $\left|\Delta\right|\simeq\left|\frac{N}{b^{\prime}}\right|\ll1$,
and the expansion is well-controlled. The non-perturbative regime
is given by $\left|\Delta\right|\ge1$, and the perturbative regime
by $\left|\Delta\right|<1$. Thus, a perturbative expansion is possible
when the electron Zeeman energy produced by the magnetic and/or Overhauser
field (provided by $N$ nuclear spins) is larger than the single maximum
hyperfine coupling constant $A$. In the rest of this section we apply
the Born approximation $\Sigma_{S}\simeq\Sigma_{S}^{(2)}$ to the
reduced self-energy, and perform the continuum limit for a large uniformly
polarized nuclear spin system. Later, we also consider higher orders.

\subsection{Born approximation}

In Born approximation, the memory kernels $\Sigma_{zz}(t),\,\Sigma_{++}(t)$
and inhomogeneous term $N_{z}(t)$ in \prettyref{eq:SzGME} and \prettyref{eq:SPlusGME}
are replaced by the forms obtained from the lowest-order self-energy,
i.e., $N_{z}(t)\rightarrow N_{z}^{(2)}(t),\,\Sigma_{zz}(t)\rightarrow\Sigma_{zz}^{(2)}(t),\,\Sigma_{++}(t)\rightarrow\Sigma_{++}^{(2)}(t)$.
In Laplace space, $\Sigma_{\uparrow\uparrow}^{(2)}(s)$, $\Sigma_{\uparrow\downarrow}^{(2)}(s)$,
and $\Sigma_{++}^{(2)}(s)$ are given for an arbitrary initial $h_{z}$
eigenstate $\ket{n}$ (see Eq. \prettyref{eq:SpecificInitialState})
in Appendix \ref{sec:Self-energy-expansion}, Eqs. \prettyref{eq:SigUpUpBornAnyn},
\prettyref{eq:SigUpDownBornAnyn}, and \prettyref{eq:SigPlusPlusBornAnyn}.
Inserting an initial state $\ket{n}$ for a large nuclear spin system
with uniform polarization gives (see Appendix \ref{sec:InitStateUnifPolarization}):

\begin{eqnarray}
\Sigma_{\uparrow\uparrow}^{(2)}(s) & = & -iNc_{+}\left[I_{+}(s-i\omega_{n})+I_{-}(s+i\omega_{n})\right],\label{eq:SigUpUpBornUniformPolarization}\\
\Sigma_{\uparrow\downarrow}^{(2)}(s) & = & iNc_{-}\left[I_{-}(s-i\omega_{n})+I_{+}(s+i\omega_{n})\right],\label{eq:SigUpDownBornUniformPolarization}\\
\Sigma_{++}^{(2)}(s) & = & -iN\left[c_{-}I_{+}(s)+c_{+}I_{-}(s)\right],\label{eq:SigPlusPlusBornUniformPolarization}\\
I_{\pm}(s) & = & \frac{1}{4N}\sum_{k}\frac{A_{k}^{2}}{s\mp i\frac{A_{k}}{2}}.\label{eq:IPlusMinusLaplaceTrans}\end{eqnarray}
 In the above, the coefficients\begin{equation}
c_{\pm}=I(I+1)-\left\langle \left\langle m(m\pm1)\right\rangle \right\rangle \label{eq:CPlusMinusDefinition}\end{equation}
 have been introduced, where $\left\langle \left\langle F(m)\right\rangle \right\rangle =\sum_{m=-I}^{I}P_{I}(m)F(m)$
for an arbitrary function $F(m)$. $P_{I}(m)$ is the probability
of finding a nuclear spin $I$ with $z$-projection $m$. The polarization
$p$ of the initial nuclear state is defined through the relation
$\left\langle \left\langle m\right\rangle \right\rangle =pI$. Without
loss of generality, in the rest of this paper $p>0$, but $b^{\prime}$
may take on positive or negative values. Assuming a uniform polarization
in the nuclear spin system, we can evaluate the nuclear Overhauser
field in terms of the initial polarization:\begin{equation}
\left[h_{z}\right]_{nn}=\sum_{i}A_{i}\left\langle \left\langle m\right\rangle \right\rangle =pIA,\end{equation}
 where we have used $\sum_{i}A_{i}=A$.

The continuum limit is performed by taking $N_{\mathrm{tot}}\rightarrow\infty$,
while $N\gg1$ is kept constant. For times $t\ll\sqrt{N}$, this allows
the replacement of sums by integrals $\sum_{k}\to\int_{0}^{\infty}dk$,
with small corrections (see Appendix \ref{sec:Continuum-limit}).
We insert the coupling constants $A_{k}$ from Eq. \prettyref{eq:AkDefinition}
into Eq. \prettyref{eq:IPlusMinusLaplaceTrans}, perform the continuum
limit and make the change of variables $x=\frac{A_{k}}{2}$ to obtain\begin{equation}
I_{\pm}(s)=\frac{d}{m}\int_{0}^{1}dx\frac{x\left|\ln x\right|^{\nu}}{s\mp ix},\,\,\,\,\,\nu=\frac{d}{m}-1.\label{eq:IPlusMinusLTCL}\end{equation}
 We use the relation $I_{\pm}(t=0)=\lim_{s\to\infty}sI_{\pm}(s)$
to obtain the initial amplitude\begin{equation}
I_{0}\equiv I_{\pm}(t=0)=\frac{d}{m}\left(\frac{1}{2}\right)^{\frac{d}{m}}\Gamma\left(\frac{d}{m}\right)\end{equation}
 for an arbitrary ratio $\frac{d}{m}$. For parabolic confinement
in two dimensions, $m=d=2$. The integral in \prettyref{eq:IPlusMinusLTCL}
can then be performed easily, which yields 

\begin{equation}
I_{\pm}(s)=s\left[\log(s\mp i)-\log(s)\right]\pm i\,\,\,\,\,(m=d=2).\label{eq:IPlusMinusLTCL2D}\end{equation}
In dimensionless units $\frac{A_{0}}{2}=1$, we find $A=\sum_{k}A_{k}\to\int dkA_{k}$,
with the coupling constants $A_{k}$ given in Eq. \prettyref{eq:AkDefinition}:
\begin{equation}
A=A_{0}N\frac{d}{m}\Gamma\left(\frac{d}{m}\right)=2N\frac{d}{m}\Gamma\left(\frac{d}{m}\right).\label{eq:HyperfineADimensionless}\end{equation}

\section{High field solution\label{sec:High-field-solution}}

In the next section, we will obtain a complete solution to the GME
within the Born approximation. This complete solution will exhibit
non-perturbative features (which can not be obtained from standard
perturbation theory), in the weakly perturbative regime for the self-energy,
which we define by $\left|\Delta\right|\lesssim1$. Here, we find
the leading behavior in the strongly perturbative (high magnetic field)
limit, defined by $\left|\Delta\right|\ll1$, or equivalently, $\left|b^{\prime}\right|\gg N$.
We do this in two ways. First, we apply standard perturbation theory,
where we encounter known difficulties\cite{khaetskii:2002a} (secular
terms that grow unbounded in time). Second, we extract the leading-order
spin dynamics from the non-Markovian remainder term in a Born-Markov
approximation performed directly on the GME. We find that the secular
terms are absent from the GME solution. We then give a brief description
of the dependence of the spin decay on the form and dimensionality
of the electron envelope wave function.

\subsection{Perturbation theory}

Applying standard time-dependent perturbation theory (see Appendix
\ref{sec:Perturbation-theory}) to lowest (second) order in $\mathcal{H}_{V}^{\prime}$,
performing the continuum limit, and expanding the result to leading
order in $\frac{1}{\omega_{n}}$, we find\begin{eqnarray}
\left\langle S_{+}\right\rangle _{t} & = & \sigma_{+}^{\mathrm{osc}}(t)+\sigma_{+}^{\mathrm{dec}}(t)+\sigma_{+}^{\mathrm{sec}}(t),\label{eq:PTSplus}\\
\left\langle S_{z}\right\rangle _{t} & = & \overline{\left\langle S_{z}\right\rangle }_{\infty}+\sigma_{z}^{\mathrm{dec}}(t),\label{eq:PTSz}\end{eqnarray}
 where \begin{eqnarray}
\sigma_{+}^{\mathrm{osc}}(t) & = & \left[1-\delta I_{0}\left(c_{+}+c_{-}\right)\right]\left\langle S_{+}\right\rangle _{0}e^{i\omega_{n}t},\label{eq:PTSplusOsc}\\
\sigma_{+}^{\mathrm{dec}}(t) & = & \delta\left[C_{+}^{+}I_{-}(t)+C_{-}^{+}I_{+}(t)\right],\label{eq:PTSplusDecay}\\
\sigma_{+}^{\mathrm{sec}}(t) & = & i\Delta I_{0}\left(c_{+}+c_{-}\right)\left\langle S_{+}\right\rangle _{0}t,\label{eq:PTSplusGrow}\end{eqnarray}
 and

\begin{eqnarray}
\overline{\left\langle S_{z}\right\rangle }_{\infty} & = & \left[1-2\delta I_{0}\left(c_{+}+c_{-}\right)\right]\left\langle S_{z}\right\rangle _{0}+2pI\delta I_{0},\label{eq:PTSzInf}\\
\sigma_{z}^{\mathrm{dec}}(t) & = & 2\delta\mathrm{Re}\left[e^{-i\omega_{n}t}\left(C_{+}^{z}I_{-}(t)+C_{-}^{z}I_{+}(t)\right)\right].\label{eq:PTSzDecay}\end{eqnarray}
 We have introduced the smallness parameter $\delta=\frac{N}{\omega_{n}^{2}}$
and the coefficients\begin{eqnarray}
C_{\pm}^{X} & = & \left\{ \begin{array}{c}
c_{\pm}\left(\left\langle S_{z}\right\rangle _{0}\pm\frac{1}{2}\right),\,\, X=z\\
c_{\pm}\left\langle S_{+}\right\rangle _{0},\,\, X=+.\end{array}\right.\label{eq:HighFieldCCoeffs}\end{eqnarray}
 $\left\langle S_{z}\right\rangle _{t}$ is the sum of a constant
contribution $\overline{\left\langle S_{z}\right\rangle }_{\infty}$
and a contribution that decays to zero $\sigma_{z}^{\mathrm{dec}}(t)$
with initial amplitude $O(\delta)$. The transverse spin $\left\langle S_{+}\right\rangle _{t}$
is the sum of an oscillating component $\sigma_{+}^{\mathrm{osc}}(t)$,
a decaying component $\sigma_{+}^{\mathrm{dec}}(t)$ with initial
amplitude $O(\delta)$, and a secular term $\sigma_{+}^{\mathrm{sec}}(t)$,
which grows unbounded (linearly) in time. At fourth order in $\mathcal{H}_{V}^{\prime}$,
$\left\langle S_{z}\right\rangle _{t}$ also contains a secular term.
These difficulties, which have been reported previously,\cite{khaetskii:2002a,khaetskii:2003a}
suggest the need for a more refined approach. In the next subsection
these problems will be resolved by working directly with the GME (in
Born approximation) to find the correct leading-order spin dynamics
for high magnetic fields.

\subsection{Non-Markovian corrections}

Markovian dynamics are commonly assumed in spin systems,\cite{klauder:1962a,desousa:2003a}
often leading to purely exponential relaxation and decoherence times
$T_{1}$, and $T_{2}$, respectively. For this reason, it is important
to understand the nature of corrections to the standard Born-Markov
approximation, and, as will be demonstrated in \prettyref{sec:Measurement}
on measurement, there are situations where the non-Markovian dynamics
are dominant and observable. 

To apply the Born-Markov approximation to $\left\langle S_{+}\right\rangle _{t}$,
we change variables $\left\langle S_{+}^{\prime\prime}\right\rangle _{t}=e^{-i\left(\omega_{n}+\widetilde{\omega}\right)t}\left\langle S_{+}\right\rangle _{t}$
in \prettyref{eq:SPlusGME} and substitute $\Sigma_{++}(t)\rightarrow\Sigma_{++}^{(2)}(t)$,
which gives:\begin{equation}
\dot{\left\langle S_{+}^{\prime\prime}\right\rangle }_{t}=-i\widetilde{\omega}\left\langle S_{+}^{\prime\prime}\right\rangle _{t}-i\int_{0}^{t}dt^{\prime}e^{-i\omega(t-t^{\prime})}\Sigma_{++}^{(2)}(t-t^{\prime})\left\langle S_{+}^{\prime\prime}\right\rangle _{t^{\prime}},\label{eq:SPlusMarkovRot}\end{equation}
where $\omega=\omega_{n}+\widetilde{\omega}$. We define the function
$\psi(t)=\int_{t}^{\infty}dt^{\prime}e^{-i\omega t^{\prime}}\Sigma_{++}^{(2)}(t^{\prime})$,
so that $\psi(0)=\Sigma_{++}^{(2)}(s=i\omega)$. We find\cite{fick:1990a}\begin{equation}
\dot{\left\langle S_{+}^{\prime\prime}\right\rangle }_{t}=-i\left(\psi(0)+\widetilde{\omega}\right)\left\langle S_{+}^{\prime\prime}\right\rangle _{t}+i\frac{d}{dt}\int_{0}^{t}dt^{\prime}\psi(t-t^{\prime})\left\langle S_{+}^{\prime\prime}\right\rangle _{t^{\prime}}.\end{equation}
The frequency shift $\widetilde{\omega}$ is chosen to satisfy $\widetilde{\omega}=-\mathrm{Re}\left[\psi(0)\right]=-\mathrm{Re}\left[\Sigma_{++}^{(2)}\left(s=i\left(\omega_{n}+\widetilde{\omega}\right)\right)\right]$
to remove the oscillating part from $\left\langle S^{\prime\prime}\right\rangle _{t}$.
When $\left|\omega\right|>1$, and after performing the continuum
limit, we find a \emph{vanishing} decay rate $\Gamma=-\mathrm{Im}\left[\Sigma_{++}^{(2)}\left(s=i\omega\right)\right]=0$,
which shows that there is no decay in the Markovian solution for $\left|\omega\right|>1$.
After integrating the resulting equation, we have\begin{equation}
\left\langle S_{+}^{\prime\prime}\right\rangle _{t}=\left\langle S_{+}^{\prime\prime}\right\rangle _{0}+R_{+}(t).\label{eq:SPlusRemainderInRotatingFrame}\end{equation}
The Markovian solution is given by $\left\langle S_{+}^{\prime\prime}\right\rangle _{t}=\left\langle S_{+}^{\prime\prime}\right\rangle _{0}$,
and the remainder term $R_{+}(t)=i\int_{0}^{t}dt^{\prime}\psi(t-t^{\prime})\left\langle S_{+}^{\prime\prime}\right\rangle _{t^{\prime}}$
gives the exact correction to the Markovian dynamics (within the Born
approximation). We rewrite the remainder term as \begin{equation}
R_{+}(t)=i\int_{0}^{t}dt^{\prime}\psi(t-t^{\prime})\left(\left\langle S_{+}^{\prime\prime}\right\rangle _{0}+R_{+}(t^{\prime})\right).\end{equation}
 Within the Born approximation, $R_{+}(t)$ is associated with a smallness
$O(\delta=\frac{N}{\omega_{n}^{2}})$ (since $\psi(t)\sim\Sigma_{++}^{(2)}(t)$),
so the above expression can be iterated to evaluate the leading-order
contribution to $R_{+}(t)$ in an asymptotic expansion for large $\omega_{n}$.
This gives \begin{equation}
R_{+}(t)\sim-\delta I_{0}(c_{+}+c_{-})\left\langle S_{+}\right\rangle _{0}+e^{-i\omega_{n}t}\sigma_{+}^{\mathrm{dec}}(t),\label{eq:SplusLargebRotFrame}\end{equation}
with $\sigma_{+}^{\mathrm{dec}}(t)$ given in Eq. \prettyref{eq:PTSplusDecay}.

Due to the inhomogeneous term $N_{z}(t)$ in \prettyref{eq:SzGME},
the $\left\langle S_{z}\right\rangle _{t}$ equation does not have
a simple convolution form, so it is not clear if a Markov approximation
for $\left\langle S_{z}\right\rangle _{t}$ is well-defined. However,
applying the same procedure that was used on $\left\langle S_{+}\right\rangle _{t}$
to determine the deviation of $\left\langle S_{z}\right\rangle _{t}$
from its initial value gives the remainder $R_{z}(t)$, to leading
order in $\frac{1}{\omega_{n}}$,\begin{equation}
R_{z}(t)\sim-2\delta I_{0}\left(c_{+}+c_{-}\right)\left\langle S_{z}\right\rangle _{0}+2pI\delta I_{0}+\sigma_{z}^{\mathrm{dec}}(t).\label{eq:SzLargeb}\end{equation}
 Here, $\sigma_{z}^{\mathrm{dec}}(t)$ is identical to the result
from standard perturbation theory, given by Eq. \prettyref{eq:PTSzDecay}.

Corrections to the Markov approximation can indeed be bounded for
all times to a negligible value by making the parameter $\delta$
sufficiently small. However, the dynamics with amplitude $O(\delta)$
are completely neglected within a Markov approximation. 

If we use $\left\langle S_{z}\right\rangle _{t}=\left\langle S_{z}\right\rangle _{0}+R_{z}(t)$
and Eq. \prettyref{eq:SPlusRemainderInRotatingFrame}, and return
to the rest frame for $\left\langle S_{+}\right\rangle _{t}$, Eqs.
\prettyref{eq:SplusLargebRotFrame} and \prettyref{eq:SzLargeb} recover
the high-field results from standard perturbation theory, given in
Eqs. \prettyref{eq:PTSplus} and \prettyref{eq:PTSz}, with one crucial
difference. The result from standard perturbation theory contains
a secular term, which is absent in the current case. Thus, by performing
an expansion of the self-energy instead of the spin operators directly,
the contributions that led to an unphysical divergence in $\left\langle S_{+}\right\rangle _{t}$
have been successfully re-summed. %
\begin{table}
\begin{tabular}{|c|c|c|c|}
\hline 
Symbol&
$A_{0}/2=1$, $\hbar=1$&
$A_{0}=A/N$&
$B_{z}=0$\tabularnewline
\hline
\hline 
$b^{\prime}$&
$b-\epsilon_{nz}$&
$g^{*}\mu_{B}B_{z}-g_{I}\mu_{N}B_{z}$&
$0$\tabularnewline
\hline 
$\omega_{n}$&
$b^{\prime}+2pIN$&
$b^{\prime}+pIA$&
$pIA$\tabularnewline
\hline 
$\Delta$&
$N/\omega_{n}$&
$A/2\omega_{n}$&
$1/2pI$\tabularnewline
\hline 
$\delta$&
$N/\omega_{n}^{2}$&
$A^{2}/4N\omega_{n}^{2}$&
$1/(2pI)^{2}N$\tabularnewline
\hline 
$c_{+}$&
$1-f_{\uparrow}$&
--&
--\tabularnewline
\hline 
$c_{-}$&
$f_{\uparrow}$&
--&
--\tabularnewline
\hline 
$\Omega_{0}$&
$\sqrt{\frac{N}{2}(c_{+}+c_{-})}$&
$\frac{A}{\hbar\sqrt{8N}}$&
$\frac{A}{\hbar\sqrt{8N}}$\tabularnewline
\hline 
$t_{\mathrm{hf}}/\tau_{\mathrm{hf}}$&
$1$&
$2N\hbar/A$&
$2N\hbar/A$\tabularnewline
\hline 
$t_{\mathrm{c}}/\tau_{\mathrm{c}}$&
$\frac{1}{\sqrt{N(1-p^{2})}}$&
$\frac{2\hbar}{A}\sqrt{\frac{N}{1-p^{2}}}$&
$\frac{2\hbar}{A}\sqrt{\frac{N}{1-p^{2}}}$\tabularnewline
\hline
\end{tabular}

\caption{\label{cap:Symbols}Some symbols used in the text. The second column
gives the value in dimensionless units, the third column gives the
value in dimension-full units assuming $A_{0}=\frac{A}{N}$, and the
fourth column gives the value of each symbol in zero magnetic field.
The values shown are: the effective applied field $b^{\prime}$, the
total effective field (applied field and Overhauser field) seen by
the electron $\omega_{n}$, the smallness parameter $\Delta$, which
determines the perturbative regime for electron spin dynamics, the
smallness parameter $\delta$, which bounds the deviation of the electron
spin from a Markovian solution, the coefficients $c_{+}$ and $c_{-}$,
in terms of the fraction of nuclear spins $I=\frac{1}{2}$ up in the
initial state $f_{\uparrow}$, the electron spin precession frequency
$\Omega_{0}$ when the resonance condition $\omega_{n}=0$ is satisfied,
the time scale $t_{\mathrm{hf}}$ for the decay of the electron spin
in the presence of an initial $h_{z}$ eigenstate of the nuclear system,
and the time scale $t_{\mathrm{c}}$ for the decay of the electron
spin in the presence of an ensemble of initial nuclear spin states
or a superposition of $h_{z}$ eigenstates at zeroth order in the
nuclear spin--electron spin flip-flop terms. }
\end{table}

\begin{table}
\begin{tabular}{|c|c|c|}
\hline 
&
GaAs&
Si:P\tabularnewline
\hline
\hline 
$A$&
$90\,\mu eV$&
$0.1\,\mu eV$\tabularnewline
\hline 
$N$&
$10^{5}$&
$10^{2}$\tabularnewline
\hline 
$B_{z}$&
$7\,\mathrm{T}$&
$0.1\,\mathrm{T}$\tabularnewline
\hline 
$p$&
$0$&
$0$\tabularnewline
\hline 
$\Delta$&
$0.25$&
$0.25$\tabularnewline
\hline 
$\delta$&
$10^{-6}$&
$10^{-3}$\tabularnewline
\hline 
$\Omega_{0}$&
$10^{8}\mathrm{s}^{-1}$&
$10^{7}\mathrm{s^{-1}}$\tabularnewline
\hline 
$\tau_{\mathrm{hf}}$&
$1\,\mu\mathrm{s}$&
$1\,\mu\mathrm{s}$\tabularnewline
\hline 
$\tau_{\mathrm{c}}$&
$5\,\mathrm{ns}$&
$100\,\mathrm{ns}$\tabularnewline
\hline
\end{tabular}

\caption{\label{cap:SymbolsNumericalValues}Sample numerical values for the
symbols listed in Table \prettyref{cap:Symbols} for a GaAs quantum
dot or an electron trapped at a donor impurity in natural Si:P.}
\end{table}

\subsection{Dependence on the wave function}

\begin{figure}
\includegraphics[%
  scale=0.35]{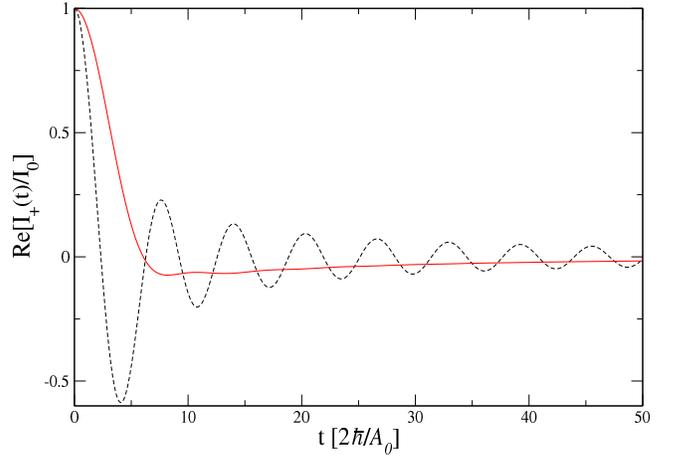}

\caption{\label{cap:WavefunctionComparison}$\mathrm{Re}\left[I_{+}(t)/I_{0}\right]$
determined numerically from Eq. \prettyref{eq:IpmTime}. For $d=3$,
$m=1$ (solid line), this corresponds to a hydrogen-like $s$-type
envelope wave function, and for $d=m=2$ (dashed line), corresponding
to a two-dimensional Gaussian envelope wave function. For the hydrogen-like
wave function, nuclear spins far from the origin, with small coupling
constants, are responsible for the slow (non-oscillatory) asymptotic
behavior. In contrast, for the Gaussian envelope wave function nuclear
spins near the center, with larger coupling constants, give rise to
oscillations in the asymptotic behavior of $I_{+}(t)/I_{0}$. }
\end{figure}

The purpose of this subsection is to evaluate the dependence of the
non-Markovian dynamics on the form of the electron envelope wave function
$\psi(r)$. The high-field dynamics, described by Eqs. \prettyref{eq:SplusLargebRotFrame}
and \prettyref{eq:SzLargeb}, depend only on the integrals $I_{\pm}(t)$.
From Eq. \prettyref{eq:IPlusMinusLTCL} we find\begin{equation}
I_{\pm}(t)=\frac{d}{m}\int_{0}^{1}dx\left|\ln x\right|^{\nu}xe^{\pm ixt},\,\,\,\,\,\nu=\frac{d}{m}-1.\label{eq:IpmTime}\end{equation}
 The time scale $\tau_{\mathrm{hf}}$ for the initial decay of $I_{\pm}(t)$
is given by the inverse bandwidth (range of integration) of the above
integral. In dimension-full units, $\tau_{\mathrm{hf}}=\frac{2\hbar}{A_{0}}$.
The long-time asymptotic behavior of $I_{\pm}(t)$ depends sensitively
on the dimensionality $d$ and the form of the envelope wave function
through the ratio $\frac{d}{m}$. When $\frac{d}{m}<2$, the major
long time contribution to \prettyref{eq:IpmTime} comes from the upper
limit $x\approx1$ corresponding to nuclear spins near the origin,
and the asymptotic form of $I_{\pm}(t)$ shows slow oscillations with
period $\frac{4\pi\hbar}{A_{0}}$:\begin{equation}
I_{\pm}(t\gg1)\propto\left(\frac{1}{t}\right)^{\frac{d}{m}}e^{\pm it},\,\,\,\,\,\frac{d}{m}<2.\label{eq:IpmOscAsymptotics}\end{equation}
 When $\frac{d}{m}\ge2$, the major contribution comes from the lower
limit $x\approx0$, i.e., nuclear spins far from the center, where
the wave function is small. The resulting decay has a slowly-varying
(non-oscillatory) envelope:\begin{equation}
I_{\pm}(t\gg1)\propto\frac{\ln^{\nu}t}{t^{2}},\,\,\,\,\,\nu=\frac{d}{m}-1\ge1.\label{eq:IpmNoOscAsymptotics}\end{equation}
 Both of the above cases can be realized in physical systems. For
an electron with an $s$-type hydrogenic wave function bound, e.g.,
to a phosphorus donor impurity in Si, $m=1$ and $d=3$, which corresponds
to the case in Eq. \prettyref{eq:IpmNoOscAsymptotics}. For an electron
trapped in a parabolic quantum dot, the envelope wave function is
a Gaussian $(m=2)$ and for $d\le3$, the asymptotics of $I_{\pm}(t)$
are described by Eq. \prettyref{eq:IpmOscAsymptotics}. These two
cases are illustrated in Fig. \prettyref{cap:WavefunctionComparison},
where $\mathrm{Re}\left[I_{+}(t)/I_{0}\right]$ is shown for $d=m=2$
and $d=3,\, m=1$. %

\section{\label{sec:Non-Markovian-dynamics}Non-Markovian dynamics}

In this section we describe a \emph{complete} calculation for the
non-Markovian electron spin dynamics within the Born approximation.
In the limit of a fully polarized initial state, our Born approximation
applied to $\left\langle S_{+}\right\rangle _{t}$ recovers the exact
solution of Ref. \onlinecite{khaetskii:2002a}. All results of this
section are, however, valid for \emph{arbitrary} polarization in high
magnetic fields when the condition $\left|\Delta\right|\ll1$ is satisfied.
In addition, we find that the remainder term is bounded by the small
parameter $\delta$, $\left|R_{X}(t)\right|\le O(\delta)$, and the
stationary limit (long-time average) of the spin can be determined
with the much weaker condition $\delta\ll1$. In zero magnetic field,
and for nuclear spin $I=\frac{1}{2}$, the relevant smallness parameter
is $\delta=\frac{1}{p^{2}N}$ (see Table \prettyref{cap:Symbols}). 

We evaluate the Laplace transforms of \prettyref{eq:SzGME}, \prettyref{eq:SPlusGME}:
$S_{X}(s)=\int_{0}^{\infty}dte^{-st}\left\langle S_{X}\right\rangle _{t},\,\mathrm{Re}[s]>0$,
$X=z,+$, to convert the integro-differential equations into a pair
of linear algebraic equations which can be solved to obtain\begin{eqnarray}
S_{z}(s) & = & \frac{\left\langle S_{z}\right\rangle _{0}+N_{z}(s)}{s+i\Sigma_{zz}(s)},\label{eq:SzLaplaceTransform}\\
S_{+}(s) & = & \frac{\left\langle S_{+}\right\rangle _{0}}{s-i\omega_{n}+i\Sigma_{++}(s)}.\label{eq:SplusLaplaceTransforms}\end{eqnarray}
 When the functions $N_{z}(s),\,\Sigma_{zz}(s),\,\Sigma_{++}(s)$
are known, the Laplace transforms in \prettyref{eq:SzLaplaceTransform}
and \prettyref{eq:SplusLaplaceTransforms} can be inverted by evaluating
the Bromwich contour integral:\begin{equation}
\left\langle S_{X}\right\rangle _{t}=\frac{1}{2\pi i}\int_{\gamma-i\infty}^{\gamma+i\infty}ds\mathrm{e}^{st}S_{X}(s),\label{eq:BromwichIntegral}\end{equation}
 where all non-analyticities of $S_{X}(s)$ lie to the left of the
line of integration. To simplify the calculation, here we specialize
to the case of an electron confined to a two-dimensional parabolic
quantum dot ($d=m=2$), where the coupling constant integrals can
be performed easily to obtain the explicit form for $I_{\pm}(s)$,
given in Eq. \prettyref{eq:IPlusMinusLTCL2D}. 

Within the Born approximation, $S_{z}(s)$ has six branch points,
located at $i\omega_{n},\, i\left(\omega_{n}\pm1\right),\,-i\omega_{n},\,-i\left(\omega_{n}\pm1\right)$.
We choose the principal branch for all logarithms, defined by $\log(z)=\ln\left|z\right|+i\arg(z)$,
where $-\pi<\arg(z)\le\pi$, in which case there are five poles in
general. Three of these poles are located on the imaginary axis and
two have finite negative real part. $S_{+}(s)$ has three branch points
(at $s=0,\,\pm i$), and three poles in general. One pole has finite
negative real part and two are located on the imaginary axis. 

\begin{figure}
\includegraphics[%
  bb=130bp 380bp 400bp 730bp,
  clip,
  scale=0.8]{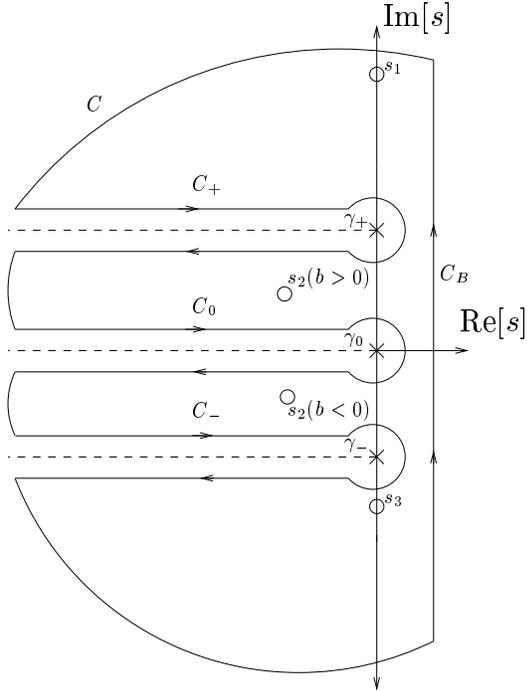}

\caption{\label{cap:Contour}The closed contour $C$ used for evaluation of
the inverse Laplace transforms of $S_{X}(s)$, $X=z,+$. All non-analyticities
of $1/D(s)$ are shown above, where $D(s)$ is given in Eq. \prettyref{eq:BornApproxDenominator}.
Branch cuts are indicated by dashed lines, branch points by crosses,
and open circles mark pole positions. The contour $C_{\alpha}$ surrounds
the branch cut extending from branch point $\gamma_{\alpha}=\alpha i$,
$\alpha=0,+,-$. When the arc that closes the contour in the negative-real
half-plane is extended to infinity, $C_{B}$ becomes the Bromwich
contour. The pole at $s_{2}$ has finite real part and is present
for $b\ne0$. The poles at $s_{1}$ and $s_{3}$ are always located
on the imaginary axis.}
\end{figure}

Applying the residue theorem to the integral around the closed contour
$C$ shown in Fig. \prettyref{cap:Contour}, $\frac{1}{2\pi i}\oint_{C}dse^{st}S_{X}(s)$,
gives\begin{eqnarray}
\left\langle S_{X}\right\rangle _{t}+\beta^{X}(t) & = & \sum_{i}P_{i}^{X}(t),\,\,\,\,\, X=z,+,\end{eqnarray}
where the pole contribution $P_{i}^{X}(t)=\mathrm{Res}\left[e^{st}S_{X}(s),\, s=s_{i}\right]$
is the residue from the pole at $s_{i}$, and the branch cut contributions
are\begin{eqnarray}
\beta^{z}(t) & = & \sum_{\alpha=0,+,-}\frac{1}{\pi}\mathrm{Im}\left[e^{-i\omega_{n}t}K_{\alpha}^{z}(t)\right],\label{eq:beta_z}\\
\beta^{+}(t) & = & \frac{1}{2\pi i}\sum_{\alpha=0,+,-}K_{\alpha}^{+}(t),\label{eq:beta_plus}\end{eqnarray}
 with branch cut integrals given by

\begin{eqnarray}
K_{\alpha}^{z}(t) & = & \int_{C_{\alpha}}dse^{st}S_{z}\left(s-i\omega_{n}\right),\\
K_{\alpha}^{+}(t) & = & \int_{C_{\alpha}}dse^{st}S_{+}\left(s\right).\end{eqnarray}
 The contour $C_{\alpha}$ runs from $\gamma_{\alpha}-\infty+i\eta$,
around $\gamma_{\alpha}$, and back to $\gamma_{\alpha}-\infty-i\eta$,
where $\eta\to0^{+}$. The branch points are given by $\gamma_{\alpha}=\alpha i,\,\alpha=0,+,-$,
as illustrated in Fig. \prettyref{cap:Contour}. In \prettyref{eq:beta_z}
we have used the fact that the branch cut integrals for $S_{z}(s)$
come in complex conjugate pairs, since $S_{z}(s^{*})=\left[S_{z}(s)\right]^{*}$.
This relationship follows directly from the definition for the Laplace
transform of the real quantity $\left\langle S_{z}\right\rangle _{t}$. 

Combining Eqs. \prettyref{eq:SigzzBA}, \prettyref{eq:SigUpUpBornUniformPolarization},
\prettyref{eq:SigUpDownBornUniformPolarization}, and \prettyref{eq:IPlusMinusLTCL2D}
to obtain $\Sigma_{zz}^{(2)}(s-i\omega_{n})$, and expanding in $\frac{1}{\omega_{n}}$
gives\begin{equation}
\Sigma_{zz}^{(2)}(s-i\omega_{n})=\Sigma_{++}^{(2)}(s)+\frac{\Delta}{4}(c_{+}+c_{-})+O\left(\delta\right),\end{equation}
 where we recall $\Delta=\frac{N}{\omega_{n}}$ and $\delta=\frac{N}{\omega_{n}^{2}}$.
The term $\frac{\Delta}{4}\left(c_{+}+c_{-}\right)$ gives rise to
a small shift in the effective magnetic field experienced by $\left\langle S_{z}\right\rangle _{t}$.
To simplify the presentation, this shift is neglected, but it could
easily be included by introducing a slight difference in the denominators
of $S_{z}(s)$ and $S_{+}(s)$. This gives \begin{eqnarray}
S_{z}(s-i\omega_{n}) & \simeq & \frac{\left\langle S_{z}\right\rangle _{0}+N_{z}^{(2)}(s-i\omega_{n})}{D(s)},\label{eq:SzBCApprox}\\
S_{+}(s) & = & \frac{\left\langle S_{+}\right\rangle _{0}}{D(s)}.\label{eq:SPlusBC}\end{eqnarray}
 The denominator $D(s)=s-i\omega_{n}+i\Sigma_{++}^{(2)}(s)$ and numerator
$N_{z}^{(2)}(s-i\omega_{n})$ are given explicitly by\begin{widetext}

\begin{eqnarray}
D(s) & = & s-ib^{\prime}+Ns\left[c_{-}\log(s-i)+c_{+}\log(s+i)-(c_{+}+c_{-})\log(s)\right],\label{eq:BornApproxDenominator}\\
N_{z}^{(2)}(s-i\omega_{n}) & = & -\frac{\Delta}{2}(c_{+}+c_{-})-i\Delta\frac{s}{2}\left[c_{+}\log(s+i)-c_{-}\log(s-i)+(c_{-}-c_{+})\log(s)\right]+O\left(\delta\right).\label{eq:BornApproxNzHighFreq}\end{eqnarray}
 \end{widetext}

The branch cuts and poles of $S_{z}(s-i\omega_{n})$ and $S_{+}(s)$,
as given in Eqs. \prettyref{eq:SzBCApprox} and \prettyref{eq:SPlusBC},
are shown in Fig. \prettyref{cap:Contour}. We note that different
analytic features will produce different types of dynamic behavior
after the inversion integral has been evaluated. The branch cut contributions
$\beta^{X}(t)$ have long-time tails that are non-exponential. Poles
with finite negative real part will give rise to exponential decay.
Poles on the imaginary axis away from the origin will lead to undamped
oscillations, and a pole at the origin will give a constant residue,
independent of time. The rest of this section is divided accordingly,
describing each type of contribution to the total time evolution of
$\left\langle S_{X}\right\rangle _{t}$.

\subsection{Non-exponential decay}

The contribution to $K_{\alpha}^{X}(t)$ circling each branch point
$\gamma_{\alpha}$ is zero, so the branch cut integrals can be rewritten
as\begin{equation}
K_{\alpha}^{X}(t)=e^{\gamma_{\alpha}t}\int_{0}^{\infty}dxe^{-xt}\xi_{X}(x,\gamma_{\alpha})\label{eq:BCIntegralExpCutoff}\end{equation}
 where \begin{multline}
\xi_{X}(x,\gamma_{\alpha})=\lim_{\eta\to0^{+}}\left[S_{X}(s_{\alpha}^{X}(x)+i\eta)\right.\\
-\left.S_{X}(s_{\alpha}^{X}(x)-i\eta)\right],\end{multline}
 with\begin{equation}
s_{\alpha}^{X}(x)=-x+\gamma_{\alpha}+\left\{ \begin{array}{c}
-i\omega_{n},\,\, X=z\\
0,\,\, X=+\end{array}\right..\end{equation}

The form of $K_{\alpha}^{X}(t)$ in Eq. \prettyref{eq:BCIntegralExpCutoff}
suggests a direct procedure for evaluating the long-time asymptotics
of the branch cut contributions. For long times, the integrand of
\prettyref{eq:BCIntegralExpCutoff} is cut off exponentially at $x\sim\frac{1}{t}\to0$.
To find the asymptotic behavior, we find the leading $x$-dependence
of $\xi_{X}(x,\gamma_{\alpha})$ for $x\to0^{+}$. We substitute this
into \prettyref{eq:BCIntegralExpCutoff}, and find the first term
in an asymptotic expansion of the remaining integral. The leading-order
long-time asymptotics obtained in this way for all branch cut integrals
$K_{\alpha}^{X}(t)$ are given explicitly in Appendix \ref{sec:Branch-cut-asymptotics}.
When $b^{\prime}=0$, the denominator $D(s)\to0$ when $s\to0$, and
the dominant asymptotic behavior comes from $K_{0}^{X}(t\to\infty)\propto\frac{1}{\ln t}$.
For $b^{\prime}\ne0$, $D(s)$ remains finite at the $s=0$ branch
point and the dominant long-time contributions come from $K_{\pm}^{X}(t\to\infty)\propto\frac{1}{t\ln^{2}t}$.
In zero magnetic field, the leading-order term in the asymptotic expansion
is dominant for times $t\gg1$, but in a finite magnetic field, the
leading term only dominates for times $t\gg e^{\left|b^{\prime}\right|/N}$.
In summary,\begin{eqnarray}
\beta^{X}(t\gg1) & \propto & \frac{1}{\ln t},\,\,\,\,\,\,\,\,\,\,\,\, b^{\prime}=0,\label{eq:BranchCutsZeroMagField}\\
\beta^{X}(t\gg e^{\left|b^{\prime}\right|/N}) & \propto & \frac{1}{t\ln^{2}t},\,\,\,\,\,\,\,\,\,\, b^{\prime}\ne0.\label{eq:BranchCutsNonzeroMagField}\end{eqnarray}
 This is in agreement with the exact result\cite{khaetskii:2003a}
for a fully-polarized system of nuclear spins $I=\frac{1}{2}$ in
a two-dimensional quantum dot. This inverse logarithmic time dependence
cannot be obtained from the high-field solutions of \prettyref{sec:High-field-solution}.
The method used here to evaluate the asymptotics of the Born approximation
therefore represents a nontrivial extension of the exact solution
to a nuclear spin system of reduced polarization, but with $\left|\Delta\right|<1$
(see Table \prettyref{cap:Symbols}). 

The branch cut integrals can be evaluated for shorter times in a way
that is asymptotically exact in a high magnetic field. To do this,
we expand the integrand of Eq. \prettyref{eq:BCIntegralExpCutoff}
to leading nontrivial order in $\frac{1}{\omega_{n}}$, taking care
to account for any singular contributions. For asymptotically large
positive magnetic fields, we find (see Appendix \ref{sec:Branch-cut-asymptotics}):\begin{multline}
\sum_{\alpha}K_{\alpha}^{X}(t)\sim-i2\pi\delta\left(C_{-}^{X}I_{+}(t)+C_{+}^{X}I_{-}(t)\right)\\
-\frac{C_{-}^{X}}{Nc_{-}^{2}}e^{-z_{0}t}\label{eq:HighFieldBCIntegrals}\end{multline}
 with coefficients $C_{\pm}^{X}$ given in \prettyref{eq:HighFieldCCoeffs}
and in the above,\begin{eqnarray}
z_{0} & = & x_{0}-i\epsilon(x_{0}),\label{eq:z0Definition}\\
x_{0} & = & \frac{\omega_{n}}{2\pi Nc_{-}},\label{eq:x0Definition}\\
\epsilon(x) & = & \frac{x}{2\pi c_{-}N}+\frac{c_{+}+c_{-}}{4\pi c_{-}x}.\label{eq:epsDefinition}\end{eqnarray}
 In high magnetic fields, we will show that the exponential contribution
to Eq. \prettyref{eq:HighFieldBCIntegrals} \emph{cancels} with the
contribution from the pole at $s_{2}$, $P_{2}^{X}(t)$. We stress
that this result is only true in the high-field limit $\frac{\left|b^{\prime}\right|}{N}\gg1$,
where the asymptotics are valid.

\subsection{Exponential decay}

\begin{figure}
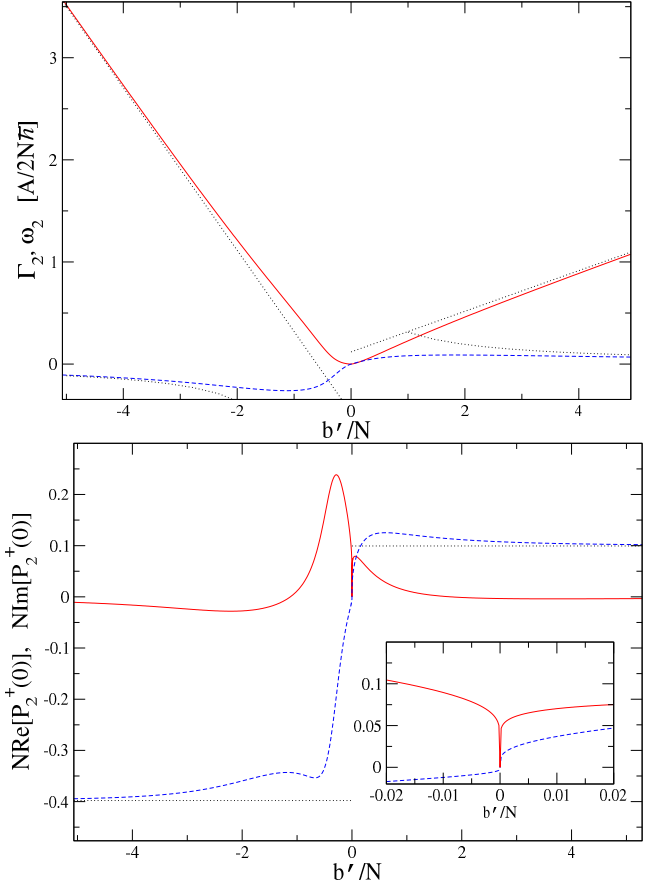

\includegraphics[%
  clip,
  scale=0.35]{fig4a.eps}

\includegraphics[%
  clip,
  scale=0.35]{fig4b.eps}

\caption{\label{cap:Pole2GammaOmega}Top: numerically determined rate $\Gamma_{2}$
(solid line) and frequency renormalization $\omega_{2}$ (dashed line)
as a function of magnetic field $b^{\prime}/N$. Bottom: $N\mathrm{Re}\left[P_{2}^{+}(0)\right]$
(solid line) and $N\mathrm{Im}\left[P_{2}^{+}(0)\right]$ (dashed
line) as a function of magnetic field for the initial state $\left\langle S_{+}\right\rangle _{0}=\left\langle S_{x}\right\rangle _{0}=\frac{1}{2}$.
The dotted lines give the asymptotics for high magnetic fields from
Eqs. \prettyref{eq:Gamma2Largeb}, \prettyref{eq:Omega2Largeb}, and
\prettyref{eq:P20Largeb}. The parameters used were $p=0.6$, $N=10^{5}$,
$I=\frac{1}{2}$. }
\end{figure}

When $b^{\prime}=0$, there are no poles with finite real part. For
$b^{\prime}\ne0$, a pole (at $s_{2}$ in Fig. \prettyref{cap:Contour})
emerges from the branch point at $s=0$. The pole contribution $P_{2}^{X}(t)$
decays exponentially with rate $\Gamma_{2}=-\mathrm{Re}[s_{2}]$,
and has an envelope that oscillates at a frequency determined by $\omega_{2}=\mathrm{Im}[s_{2}]$:\begin{eqnarray}
P_{2}^{z}(t) & = & e^{-\Gamma_{2}t}e^{-i(\omega_{n}-\omega_{2})t}P_{2}^{z}(0)\\
P_{2}^{+}(t) & = & e^{-\Gamma_{2}t}e^{i\omega_{2}t}P_{2}^{+}(0).\end{eqnarray}
 Setting $s_{2}=-\Gamma_{2}+i\omega_{2}$, we find the decay rate
$\Gamma_{2}$, frequency renormalization $\omega_{2}$, and amplitudes
of these pole contributions from asymptotic solutions to the pair
of equations $\mathrm{Re}[D(s_{2})]=\mathrm{Im}[D(s_{2})]=0$ and
$P_{2}^{X}(0)=\mathrm{Res[S_{X}(s)},s=s_{2}]$ for high and low magnetic
fields $b^{\prime}$. $\Gamma_{2},\,\omega_{2}$, and $P_{2}^{X}(0)$
have the asymptotic field dependences (for high magnetic fields $b^{\prime}\gg N$):\begin{eqnarray}
\Gamma_{2} & \sim & \pm\frac{\omega_{n}}{2\pi Nc_{\mp}},\,\,\,\,\omega_{n}\gtrless0,\label{eq:Gamma2Largeb}\\
\omega_{2} & \sim & \pm\frac{\Gamma_{2}}{2\pi c_{\mp}N}\pm\frac{c_{+}+c_{-}}{4\pi c_{\mp}\Gamma_{2}},\,\,\,\,\omega_{n}\gtrless0,\label{eq:Omega2Largeb}\\
P_{2}^{X}(0) & \sim & \frac{C_{\mp}^{X}/c_{\mp}}{1\mp i2\pi Nc_{\mp}},\,\,\,\,\omega_{n}\gtrless0.\label{eq:P20Largeb}\end{eqnarray}
 Although it does not correspond to the perturbative regime, it is
interesting to consider the behavior of the exponentially decaying
pole contribution $P_{2}^{X}(t)$ in the limit $b^{\prime}\to0$,
since the Hamiltonian $\mathcal{H}$ in Eq. \prettyref{eq:HFHamiltonian}
is known to be integrable for $B_{z}=0$ ($b^{\prime}=0$).\cite{schliemann:2003a}
For vanishing positive magnetic fields $\left(b^{\prime}\to0^{+}\right)$,
with logarithmic corrections in $\frac{b^{\prime}}{Eb_{0}}$, where
$b_{0}=N(c_{+}+c_{-})$ and $E=\exp\left\{ 1+O\left(\frac{1}{N}\right)\right\} $:\begin{eqnarray}
\Gamma_{2} & \sim & \frac{\zeta b^{\prime}/b_{0}}{\ln^{2}\left(\frac{b^{\prime}}{Eb_{0}}\right)},\\
\omega_{2} & \sim & -\frac{b^{\prime}/b_{0}}{\ln\left(\frac{b^{\prime}}{Eb_{0}}\right)},\\
P_{2}^{+}(0) & \sim & -\frac{\left\langle S_{+}\right\rangle _{0}}{N(c_{+}+c_{-})\ln\left(\frac{b^{\prime}}{b_{0}}\right)},\\
P_{2}^{z}(0) & \sim & -\frac{\left\langle S_{z}\right\rangle _{0}-\left(c_{+}+c_{-}\right)/2pI}{N(c_{+}+c_{-})\ln\left(\frac{b^{\prime}}{b_{0}}\right)},\end{eqnarray}
 where $\zeta=\frac{\pi c_{-}}{c_{+}+c_{-}}$. The exponentially decaying
contribution vanishes only when $b^{\prime}=0$, and does so in an
interval that is logarithmically narrow. We have determined the rate,
frequency renormalization, and amplitude of the pole contribution
$P_{2}^{+}(t)$ numerically. The results are given in Fig. \prettyref{cap:Pole2GammaOmega}
along with the above asymptotics for high magnetic fields, $\left|b^{\prime}\right|\gg N$.

\subsection{Undamped oscillations}

The point $s_{1}$ in Fig. \prettyref{cap:Contour} corresponds to
$s=0$ for $S_{z}(s)$, so undamped oscillations in $\left\langle S_{z}\right\rangle _{t}$
arise only from the pole at $s_{3}$:\begin{equation}
P_{3}^{z}(t)=e^{-i(\omega_{n}-\omega_{3})t}P_{3}^{z}(0).\end{equation}
 Both poles on the imaginary axis give undamped oscillations in $\left\langle S_{+}\right\rangle $:\begin{equation}
P_{1}^{+}(t)+P_{3}^{+}(t)=e^{i\omega_{1}t}P_{1}^{+}(0)+e^{i\omega_{3}t}P_{3}^{+}(0).\end{equation}
 For high magnetic fields, $\left|b^{\prime}/N\right|\gg1$, \begin{eqnarray}
\omega_{1/3} & \sim & b^{\prime}+2pIN=\omega_{n},\,\,\,\,\, b^{\prime}\gtrless0,\label{eq:PrecessionPole}\\
\omega_{3/1} & \sim & \mp1\mp f_{\pm}\exp\left(-\frac{\left|b^{\prime}\right|}{c_{\pm}N}\right),\,\,\,\,\, b^{\prime}\gtrless0,\label{eq:NuclearPrecessionPole}\end{eqnarray}
 where $f_{\pm}=\left(\frac{1}{2}\right)^{\left(\frac{c_{\mp}}{c_{\pm}}\right)}\left(1+O\left(\frac{1}{N}\right)\right)$.
The frequency in Eq. \prettyref{eq:PrecessionPole} corresponds to
a simple precession of the electron spin in the sum of the magnetic
and Overhauser fields. The second frequency, Eq. \prettyref{eq:NuclearPrecessionPole},
describes the back-action of the electron spin, in response to the
slow precession of the nuclear spins in the effective field of the
electron.

For large $b^{\prime}$, the pole corresponding to simple precession
is dominant, while the other has a residue that vanishes exponentially: 

\begin{eqnarray}
P_{1/3}^{+}(0) & \sim & \frac{\left\langle S_{+}\right\rangle _{0}}{1+\frac{1}{2}(c_{+}+c_{-})\delta},\,\,\,\,\, b^{\prime}\gtrless0,\label{eq:PPlusDominantResidue}\\
P_{3/1}^{+}(0) & \sim & \frac{\left\langle S_{+}\right\rangle _{0}}{Nc_{\pm}}f_{\pm}\exp\left(-\frac{\left|b^{\prime}\right|}{c_{\pm}N}\right),\,\,\,\,\, b^{\prime}\gtrless0,\\
P_{3}^{z}(0) & \sim & \frac{b^{\prime}}{2c_{+}N}f_{+}\exp\left(-\frac{\left|b^{\prime}\right|}{c_{+}N}\right),\,\,\,\,\, b^{\prime}>0.\end{eqnarray}
 When the magnetic field $b^{\prime}$ compensates the nuclear Overhauser
field $\left[h_{z}\right]_{nn}$ ($\omega_{n}\approx0$, the usual
ESR resonance condition in the rotating frame), the poles at points
$s_{1}$ and $s_{3}$ have equal weight, and are the dominant contribution
to the electron spin dynamics. Since the resonance condition corresponds
to the strongly non-perturbative regime, $\left|\Delta\right|\gg1$,
we delay a detailed discussion of the resonance until \prettyref{sec:Beyond-Born}. 

\begin{figure}
\includegraphics[%
  scale=0.3]{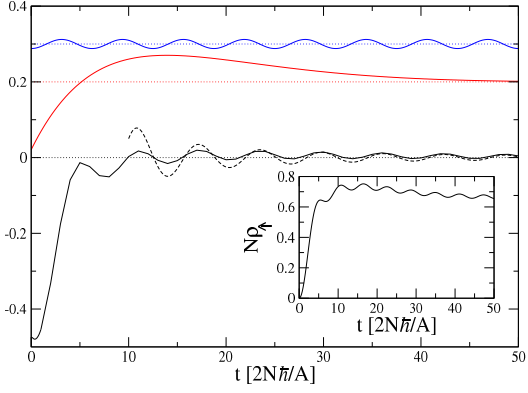}

\caption{\label{cap:WeakFieldContributions}Contributions to the inverse Laplace
transform of $\left\langle S_{z}\right\rangle _{t}$. We show the
envelopes of the rapidly oscillating functions $2N\mathrm{Re}\left[P_{3}^{z}(t)\right]+0.3$,
$2N\mathrm{Re}\left[P_{2}^{z}(t)\right]+0.2$, and $-N\beta^{z}(t)$,
determined numerically. The long-time asymptotics of $\beta^{z}(t)$
from Appendix \ref{sec:Branch-cut-asymptotics} are also shown (dashed
line). The sum of all contributions is used to obtain the population
of the spin-up state: $\rho_{\uparrow}(t)=\frac{1}{2}+\left\langle S_{z}\right\rangle _{t}$
(inset). The electron spin begins down: $\left\langle S_{z}\right\rangle _{0}=-\frac{1}{2}$.
Other parameters were $I=\frac{1}{2}$, $N=10^{5}$, $b^{\prime}=\frac{N}{2}$
(this value of $b^{\prime}$ gives, e.g., $B_{z}\simeq1\mathrm{T}$
in GaAs), and $p=0.6$. The time $t$ is given in units of $\frac{2\hbar}{A_{0}}=\frac{2N\hbar}{A}$
for $d=m=2$ in Eq. \prettyref{eq:HyperfineADimensionless} ($\frac{2N\hbar}{A}\simeq1\mathrm{\mu s}$
in GaAs). These values correspond to the weakly perturbative regime,
with $\Delta=\frac{10}{11}<1$. Note that $\rho_{\uparrow}(t)\lesssim\frac{1}{N}$
for all times.}
\end{figure}

\subsection{Stationary limit}

The contribution to $\left\langle S_{z}\right\rangle _{t}$ from the
pole at $s=0$ gives the long-time average value $\overline{\left\langle S_{z}\right\rangle }_{\infty}$,
which we define as the stationary limit:\begin{equation}
\overline{\left\langle S_{z}\right\rangle }_{\infty}=\lim_{T\to\infty}\frac{1}{T}\int_{0}^{T}\left\langle S_{z}\right\rangle _{t}dt=\lim_{s\to0}sS_{z}(s).\end{equation}
 Within the Born approximation, we find\begin{equation}
\overline{\left\langle S_{z}\right\rangle }_{\infty}=\frac{\left\langle S_{z}\right\rangle _{0}+pI\delta+O\left(\frac{N}{\omega_{n}^{4}}\right)}{1+(c_{+}+c_{-})\delta+O\left(\frac{N}{\omega_{n}^{4}}\right)}.\label{eq:SzStatLimit}\end{equation}
 The result in Eq. \prettyref{eq:SzStatLimit} follows from Eqs. \prettyref{eq:SzLaplaceTransform},
\prettyref{eq:NzBA}, \prettyref{eq:SigzzBA}, \prettyref{eq:SigUpUpBornUniformPolarization},
\prettyref{eq:SigUpDownBornUniformPolarization}, and \prettyref{eq:IPlusMinusLaplaceTrans}
by expanding the numerator and denominator in $\frac{1}{\omega_{n}}$,
using the coupling constants $A_{k}=2e^{-k/N}$ and performing the
continuum limit. $\overline{\left\langle S_{z}\right\rangle }_{\infty}$
gives the stationary level populations for spin-up and spin-down:
$\bar{\rho}_{\uparrow/\downarrow}=\frac{1}{2}\pm\overline{\left\langle S_{z}\right\rangle }_{\infty}$,
which would be fixed by the initial conditions in the absence of the
hyperfine interaction. This difference in $\bar{\rho}_{\uparrow/\downarrow}$
from the initial values can be regarded as leakage due to the nuclear
spin environment. We note that the stationary value depends on the
initial value $\left\langle S_{z}\right\rangle _{0}$, from which
it deviates only by a small amount of order $\delta$. This means,
in particular, that the system is non-ergodic. We will find that corrections
to $\overline{\left\langle S_{z}\right\rangle }_{\infty}$ at fourth
order in the flip-flop terms will be of order $\delta^{2}$, so that
the stationary limit can be determined even outside of the perturbative
regime $\left|\Delta\right|<1$, in zero magnetic field, where $\delta=\frac{1}{p^{2}N}$
for $I=\frac{1}{2}$, provided $p\gg\frac{1}{\sqrt{N}}$.

\subsection{Summary}

The results of this section for low magnetic fields are summarized
in Fig. \prettyref{cap:WeakFieldContributions}, which corresponds
to the weakly perturbative case, $\left|\Delta\right|\lesssim1$,
and displays all of the dynamical features outlined here. 

\begin{figure}
\includegraphics[%
  scale=0.35]{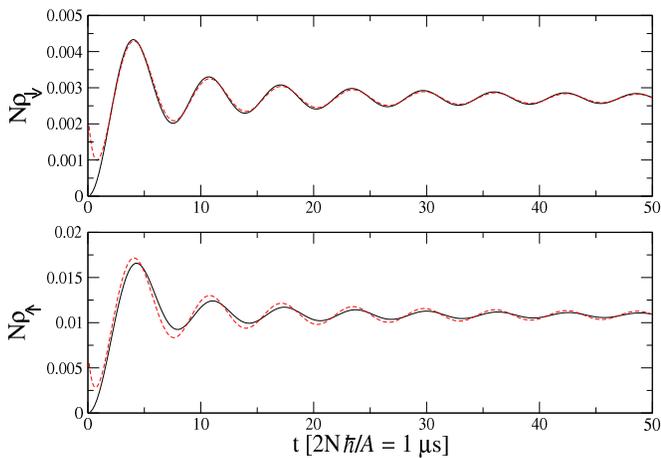}

\caption{\label{cap:HighFieldContributions}Envelope of the time-dependent
spin level populations in high magnetic fields. We give results from
numerical inversion of the Laplace transform (solid line) and the
asymptotic branch cut integral for high magnetic fields combined with
numerical results for the pole positions and residues (dashed line).
Top: spin-down level population when the electron begins in the up
state, along the nuclear spin polarization direction ($\left\langle S_{z}\right\rangle _{0}=\frac{1}{2}$).
Bottom: spin-up population for an electron that begins pointing in
the opposite direction ($\left\langle S_{z}\right\rangle _{0}=-\frac{1}{2}$).
The parameters used were $N=10^{5}$, $p=0.6$, $I=\frac{1}{2}$ and
$b^{\prime}=8N$, corresponding to a field of $B_{z}=14\,\mathrm{T}$
in GaAs. }
\end{figure}

In very high magnetic fields $\left(b^{\prime}\gg N\right)$, corresponding
to the strongly perturbative case, we combine Eqs. \prettyref{eq:HighFieldBCIntegrals},
\prettyref{eq:PrecessionPole}, \prettyref{eq:PPlusDominantResidue},
and \prettyref{eq:SzStatLimit} to obtain the asymptotic forms to
leading order in $\frac{1}{\omega_{n}}$: \begin{eqnarray}
\left\langle S_{+}\right\rangle _{t} & \sim & \sigma_{+}^{\mathrm{osc}}(t)+\sigma_{+}^{\mathrm{dec}}(t),\\
\left\langle S_{z}\right\rangle _{t} & \sim & \overline{\left\langle S_{z}\right\rangle }_{\infty}+\sigma_{z}^{\mathrm{dec}}(t),\end{eqnarray}
 where the functions $\sigma_{+}^{\mathrm{osc}}(t),\,\sigma_{+}^{\mathrm{dec}}(t),\,\overline{\left\langle S_{z}\right\rangle }_{\infty},\,\sigma_{z}^{\mathrm{dec}}(t)$,
given in Eqs. \prettyref{eq:PTSplusOsc}, \prettyref{eq:PTSplusDecay},\prettyref{eq:PTSzInf},
and \prettyref{eq:PTSzDecay} are evaluated for $d=m=2$. We stress
that $\sigma_{X}^{\mathrm{dec}}(t)\propto\delta\ll1$ is a small fraction
of the total spin. The exponentially decaying contribution from $P_{2}^{X}(t)$
is \emph{canceled} by the exponential part of the high-field branch
cut, given in Eq. \prettyref{eq:HighFieldBCIntegrals}. This result
is in agreement with the high-field asymptotic forms found earlier
in \prettyref{sec:High-field-solution}. Numerical results for the
level populations $\rho_{\uparrow/\downarrow}(t)=\frac{1}{2}\pm\left\langle S_{z}\right\rangle _{t}$
are given in Fig. \prettyref{cap:HighFieldContributions} along with
the above asymptotic forms. The secular term that appeared at lowest
order in the standard perturbation expansion of $\left\langle S_{+}\right\rangle _{t}$
is again absent from the result obtained here via the GME. At fourth
order, $t$-linear terms also appear in the standard perturbation
expansion for the longitudinal spin $\left\langle S_{z}\right\rangle _{t}$.\cite{khaetskii:2002a,khaetskii:2003a}
Due to the numerator term $N_{z}(s)$ in the expression for $S_{z}(s)$
(Eq. \prettyref{eq:SzLaplaceTransform}), it is not clear if all divergences
have been re-summed for $\left\langle S_{z}\right\rangle _{t}$ in
the perturbative expansion of the self-energy. This question is addressed
in Sec. \ref{sec:Beyond-Born} with an explicit calculation of the
fourth-order spin dynamics. 

In the next section we propose a method that could be used to probe
the non-Markovian electron spin dynamics experimentally.%

\section{\label{sec:Measurement}Measurement}

In high magnetic fields $(b^{\prime}\gg N)$, the decaying fraction
of the electron spin is very small $\left(O\left(\delta\approx\frac{N}{b^{\prime2}}\right)\right)$.
Nevertheless, the large separation between the hyperfine interaction
decay time ($\tau_{\mathrm{hf}}=\frac{2\hbar}{A_{0}}\approx1\mathrm{\mu s}$)
and the dipolar correlation time ($\tau_{\mathrm{dd}}\approx100\,\mathrm{\mu s}$
in GaAs) of the nuclear spins should allow one to obtain valuable
information about the electron spin decay from a conventional spin
echo technique applied to an ensemble of electron spins. 

In principle, the non-Markovian electron spin dynamics should be visible
in the electron spin echo envelope obtained by applying the conventional
Hahn echo sequence:\cite{slichter:1980a} $\frac{\pi}{2}-\tau-\pi_{x}-\tau-\mathrm{ECHO}$
to a large ensemble of electron spins. This can be done by conventional
means for an electron trapped at donor impurities in a solid,\cite{tyryshkin:2003a}
or from a measurement of transport current through a quantum dot.\cite{engel:2001a,engel:2002a}
The effect of this echo sequence can be summarized as follows. The
electron spins are initially aligned along the external magnetic field
$B_{z}$. At time $t=0$ the spins are tipped into the $x-y$ plane
with an initial $\frac{\pi}{2}$-pulse. Each spin precesses in its
own local effective magnetic field $\omega_{n}$. The phase factor
$e^{i\omega_{n}t}$ winds in the {}``forward'' direction for a time
$\tau$. The sign of $\omega_{n}$ (direction of the local magnetic
field) is then effectively reversed with a $\pi$-pulse along the
$x$-axis: $\omega_{n}\to-\omega_{n}$. The phase factor $e^{-i\omega_{n}t}$
unwinds in the following time interval $\tau$, and the electron spin
magnetization refocuses to give an echo when the phase factor $e^{-i\omega_{n}2\tau}=1$
simultaneously for all spins in the ensemble. As is usually assumed,
we take the pulse times and measurement time during the echo to be
negligible.\cite{slichter:1980a} The spin echo envelope gives the
ensemble magnetization (the electron spin expectation value) at the
time of the echo as a function of the free evolution time $2\tau$
before the echo. We note that the decaying fraction of $\left\langle S_{z}\right\rangle _{t}$,
$\sigma_{z}^{\mathrm{dec}}(t)$, also precesses with the phase factor
$e^{i\omega_{n}t}$ (see Eq. \prettyref{eq:PTSzDecay}), so the same
pulse sequence can also be applied to measure the decay of the longitudinal
spin, omitting the initial $\frac{\pi}{2}$-pulse. The Hahn echo envelope
should show a small initial decay by $O(\delta)$ in a time scale
$\tau_{\mathrm{hf}}$ due to the contact hyperfine interaction, followed
by a slow decay due to spectral diffusion\cite{desousa:2003a,desousa:2003b,abe:2004a}
with a time scale $\tau_{\mathrm{dd}}\approx10^{-4}\,\mathrm{s}$.
We note that a rapid initial decay of the Hahn echo envelope has been
measured for natural $\mathrm{Si:P}$, but is absent in isotopically
enriched $\mathrm{^{28}Si:P}$, in which no nuclei carry spin.\cite{fanciulli:2003a}

The fraction of the spin that decays in the time $\tau_{\mathrm{hf}}$
is small, of order $\delta$, in the perturbative regime. It may be
difficult to detect this small fraction using the conventional Hahn
echo. This problem can be reduced by taking advantage of the quantum
Zeno effect, using the Carr-Purcell-Meiboom-Gill (CPMG) echo sequence
$\frac{\pi}{2}-\left(\tau-\pi_{x}-\tau-\mathrm{ECHO}-\tau-\pi_{-x}-\tau-\mathrm{ECHO}\right)_{\mathrm{repeat}}$.
During each free evolution time between echoes, the electron decays
by an amount of order $\delta$. At each echo, a measurement of the
electron spin magnetization is performed. For a large ensemble of
electron spins, this measurement determines the state $\rho_{S}$
of the electron spin ensemble, forcing the total system into a direct
product of electron and nuclear states, as in Eq. \prettyref{eq:SuddenApproximationProductState}.
Repetition of such measurement cycles will then reveal the spin decay
due to the hyperfine interaction (by order $\delta$ after each measurement)
until the magnetization envelope reaches its stationary value. If
the electron spin decays during the free evolution time due to spectral
diffusion with a Gaussian envelope, then we require the condition
$\left(\frac{2\tau}{\tau_{dd}}\right)^{2}\ll\delta\ll1$ for the effect
of spectral diffusion to be negligible compared to the effect of the
hyperfine interaction.%
\footnote{Abe \emph{et al.}\cite{abe:2004a} have recently measured a pure Gaussian
decay of the Hahn spin echo envelope with time scale given by the
dipolar correlation time $\tau_{\mathrm{dd}}\approx10^{-4}\,\mathrm{s}$
for electrons trapped at phosphorus donors in isotopically enriched
$\mathrm{^{29}Si:P}$, where all silicon nuclei carry spin $I=\frac{1}{2}$.
In contrast to the CPMG echo sequence, only a single measurement (a
single echo) is made following each preparation in the Hahn technique.
We assume the echo envelope is the product of a Gaussian with time
scale $\tau_{\mathrm{dd}}$ and a part $f(2\tau)=1-O(\delta),\,2\tau\gtrsim\tau_{\mathrm{hf}}$,
that gives the decay due to the contact hyperfine interaction: $\exp\left[-\frac{1}{2}\left(\frac{2\tau}{\tau_{dd}}\right)^{2}\right]f(2\tau)\approx1-\frac{1}{2}\left(\frac{2\tau}{\tau_{dd}}\right)^{2}-O(\delta)$,
for times $\tau_{\mathrm{hf}}\lesssim2\tau\ll\tau_{\mathrm{dd}}$.
When $\left(\frac{2\tau}{\tau_{dd}}\right)^{2}\ll\delta$, the dominant
contribution comes from $f(2\tau)$ at each echo of the CPMG sequence. %
} The non-Markovian remainder term gives the total change in electron
spin that has occurred during the free evolution time $2\tau$: $\left.R_{X}(2\tau)\right|_{e^{\pm i\omega_{n}2\tau}=1}=\left.\left\langle S_{X}\right\rangle _{2\tau}-\left\langle S_{X}\right\rangle _{0}\right|_{e^{\pm i\omega_{n}2\tau}=1}=M_{X}(2\tau)-M_{X}(0)$,
where $M_{X}(t)$ is the CPMG magnetization envelope. In high magnetic
fields, and when there are many echoes before the magnetization envelope
decays, the CPMG magnetization envelopes $M_{X}(t)$ will therefore
obey the differential equations\begin{equation}
\frac{d}{dt}M_{X}(t)=\left.\frac{R_{X}(2\tau)}{2\tau}\right|_{\left\langle S_{X}\right\rangle _{0}=M_{X}(t),\, e^{\pm i\omega_{n}2\tau}=1},\,\,\,\,\, X=+,z,\label{eq:MXDE}\end{equation}
 where the high-field expressions for $R_{X}(t)$, given in Eqs. \prettyref{eq:SplusLargebRotFrame}
and \prettyref{eq:SzLargeb}, should be used. Thus, the decay rate
of the CPMG echo envelope $M_{X}$, as a function of the free evolution
time $2\tau$, is a \emph{direct probe} of the non-Markovian remainder
term $R_{X}(t)$. 

\begin{figure}
\includegraphics[%
  scale=0.35]{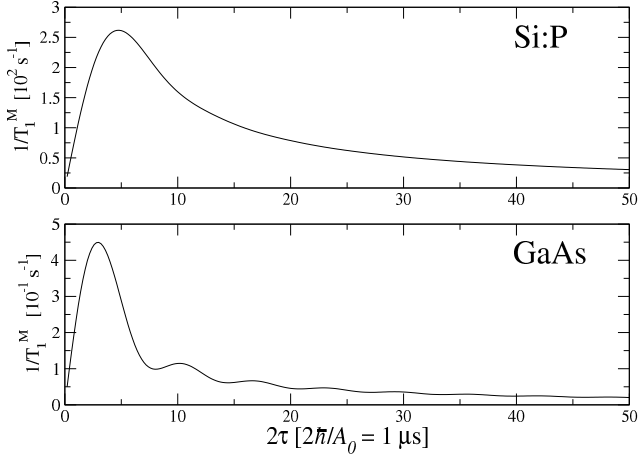}

\caption{\label{cap:DecayRates}Longitudinal decay rate $\frac{1}{T_{1}^{M}}$
of the CPMG echo envelope as a function of the free evolution time
$2\tau$ between $\pi$-pulses for an electron trapped at a phosphorus
donor impurity in Si:P (top) and in a two-dimensional GaAs quantum
dot (bottom). The free evolution time is given in units of $\frac{2\hbar}{A_{0}}\approx\frac{2N\hbar}{A}$
(the equality is exact for $d=m=2$ in Eq. \prettyref{eq:HyperfineADimensionless}).
In a GaAs quantum dot containing $N=10^{5}$ nuclei or for an electron
trapped at a shallow donor impurity in Si:P with $N=100$ nuclear
spins within one Bohr radius, $\frac{2N\hbar}{A}\approx1\,\mu\mathrm{s}$.
We have used $I=\frac{1}{2}$, $p=0.6$, and magnetic field values
from Table \prettyref{cap:SymbolsNumericalValues} to determine the
frequency units on the vertical axis. }
\end{figure}

Since the magnetization envelopes $M_{X}(t)$ are found as the result
of an ensemble measurement, it is necessary to perform an average
over different nuclear initial states $\ket{n}$ that may enter into
the solutions to Eq. \prettyref{eq:MXDE}. The local field-dependent
phase factors have been removed by the echo sequence, so the only
effect of the ensemble average is to average over $\delta=\frac{N}{\omega_{n}^{2}}$
and $c_{\pm}$, which appear in the overall amplitude of $\frac{d}{dt}M_{X}(t)$.
The relative fluctuations in these quantities are always suppressed
by the factor $\frac{1}{\sqrt{N}}$ for a large nuclear spin system. 

In the high-field limit, we find the longitudinal and transverse magnetization
envelopes $M_{z}(t)$ and $M_{+}(t)$ decay exponentially with time
constants $T_{1}^{M}$ and $T_{2}^{M}=2T_{1}^{M}$, respectively.
$M_{+}(t)$ decays to zero, and $M_{z}(t)$ decays to the limiting
value\[
M_{z}(\infty)=\frac{1}{2}\frac{c_{-}-c_{+}}{c_{-}+c_{+}}=\frac{pI}{c_{-}+c_{+}}.\]
 For nuclear spin $I=\frac{1}{2}$, $M_{z}(\infty)=\frac{p}{2}$,
i.e., the electron magnetization acquires the polarization of the
nuclear spin bath. However, since $c_{\pm}\propto I^{2}$, $M_{z}(\infty)\to0$
in the large-spin limit. Thus, a \emph{larger} fraction of the electron
spin decays in the limit of large nuclear spin. We give plots of the
longitudinal spin decay rate for $M_{z}(t)$, $\frac{1}{T_{1}^{M}}$,
as a function of the free evolution time $2\tau$ for two types of
envelope wave function in Fig. \ref{cap:DecayRates}. These plots
have been determined by integrating Eq. \prettyref{eq:MXDE} using
the high-field expression for $R_{z}(t)$ given in Eq. \prettyref{eq:SzLargeb}.
No ensemble averaging has been performed to generate these plots.
When $2\tau\ll\tau_{\mathrm{hf}}$, the envelope decay rate increases
as a function of $2\tau$ as more of the electron spin is allowed
to decay before each measurement. The rates reach a maximum at some
time $2\tau\approx\tau_{\mathrm{hf}}$, and for $2\tau\gg\tau_{\mathrm{hf}}$,
the electron spin saturates at its stationary value and the envelope
decay rates $\propto\frac{1}{2\tau}$ are determined only by the free
evolution time. Note that there are slow oscillations in the CPMG
decay rate for an electron in a GaAs quantum dot, with a Gaussian
wave function, but none for an electron trapped at a donor impurity
in Si:P.

\section{Beyond Born\label{sec:Beyond-Born}}

The goal of this section is to address the range of validity of the
results obtained in Sec. \ref{sec:Non-Markovian-dynamics}. First,
we show that the Born approximation for $\left\langle S_{+}\right\rangle _{t}$
recovers the exact solution for $I=\frac{1}{2}$, $p=1$. We then
discuss the behavior of the Born approximation near the ESR resonance,
where $\omega_{n}\approx0$. Finally, we consider the expression for
$\left\langle S_{z}\right\rangle _{t}$, obtained by including all
fourth-order corrections to the reduced self-energy, and show that
our expression is well-behaved in the continuum limit.

\subsection{Recovery of the exact solution}

When $I=\frac{1}{2}$ and $p=1$, we have $c_{-}=1$ and $c_{+}=0$,
which gives $\Sigma_{++}^{(2)}(s)=-\frac{i}{4}\sum_{k}\frac{A_{k}^{2}}{s-i\frac{A_{k}}{2}}$
from Eq. \prettyref{eq:SigPlusPlusBornAnyn}. We insert this into
\prettyref{eq:SplusLaplaceTransforms} and use $\omega_{n}=b^{\prime}+\frac{1}{2}\sum_{k}A_{k}=b^{\prime}+\frac{A}{2}$
to obtain\begin{equation}
S_{+}(s)=\frac{\left\langle S_{+}\right\rangle _{0}}{s-i\left(b^{\prime}+\frac{A}{2}\right)+\frac{1}{4}\sum_{k}\frac{A_{k}^{2}}{s-iA_{k}/2}}.\label{eq:SPlusLTFullyPolarized}\end{equation}
The Schrödinger equation for a state of the form $\ket{\psi(t)}=\alpha_{\Uparrow}(t)\ket{\Uparrow\uparrow\uparrow\cdots}+\alpha_{\Downarrow}(t)\ket{\Downarrow\uparrow\uparrow\cdots}+\sum_{k}\beta_{k}(t)\ket{\Uparrow\uparrow\cdots\downarrow_{k}\uparrow\cdots}$,
where the large arrow gives the state of the electron spin and the
thin arrows give the states of the nuclear spins, has been written
and solved (for a fully polarized nuclear spin initial state, $\beta_{k}(t=0)=0\,\forall\, k$)
in Laplace space to find the long-time asymptotic electron spin dynamics
previously.\cite{khaetskii:2003a} In Ref. \onlinecite{khaetskii:2003a}
the symbol $\alpha(t)$ was used in place of $\alpha_{\Downarrow}(t)$.
The fully-polarized state $\ket{\Uparrow\uparrow\uparrow\cdots}$
is an eigenstate of the full Hamiltonian $\mathcal{H}^{\prime}$,
so $\alpha_{\Uparrow}(t)=e^{-\frac{i}{2}\left(b^{\prime}+\frac{A}{2}\right)t}\alpha_{\Uparrow}(0)$,
which allows us to write $S_{+}(s)=\alpha_{\Uparrow}^{*}(t=0)\alpha_{\Downarrow}\left(s-\frac{i}{2}\left(b^{\prime}+\frac{A}{2}\right)\right)$.
We solve the time-dependent Schrödinger equation for $\ket{\psi(t)}$
in Laplace space, giving \begin{equation}
\alpha_{\Downarrow}\left(s^{\prime}\right)=\frac{\alpha_{\Downarrow}(t=0^{+})}{s-i\left(b^{\prime}+\frac{A}{2}\right)+\frac{1}{4}\sum_{k}\frac{A_{k}^{2}}{s-iA_{k}/2}},\label{eq:alphaLTFullyPolarized}\end{equation}
 where $s^{\prime}=s-\frac{i}{2}\left(b^{\prime}+\frac{A}{2}\right)$.
Thus, in the limit of full polarization of the nuclear system, the
Born approximation applied to $\left\langle S_{+}\right\rangle _{t}$
becomes exact. For a fully polarized nuclear spin system $\left\langle S_{z}\right\rangle _{t}$
is given by the relationship $\left\langle S_{z}\right\rangle _{t}=\frac{1}{2}\left(1-2\left|\alpha_{\Downarrow}(t)\right|\right)=\frac{1}{2}\left(1-2\left|\frac{\left\langle S_{+}\right\rangle _{t}}{\alpha_{\Uparrow}^{*}(t=0)}\right|^{2}\right)$.
Unfortunately, this result is not recovered directly from the Born
approximation for $\left\langle S_{z}\right\rangle _{t}$, as we will
show in the next subsection.

\subsection{Resonance}

On resonance, $\omega_{n}=0$, i.e., the external field $b^{\prime}$
compensates the Overhauser field $\left[h_{z}\right]_{nn}$. The resonance
is well outside of the perturbative regime, defined by $\left|\Delta\right|=\left|\frac{N}{\omega_{n}}\right|<1$,
but we proceed in the hope that the Born approximation applied to
the self-energy captures some of the correct behavior in the non-perturbative
limit. On resonance, the major contributions to $\left\langle S_{z}\right\rangle _{t}$
come from three poles, at $s=0$, $s=s_{3}$, and $s=s_{3}^{*}$:\begin{equation}
\left\langle S_{z}\right\rangle _{t}\approx\overline{\left\langle S_{z}\right\rangle }_{\infty}+2\mathrm{Re}\left[P_{3}^{z}(t)\right].\end{equation}
\begin{figure}
\includegraphics[%
  scale=0.35]{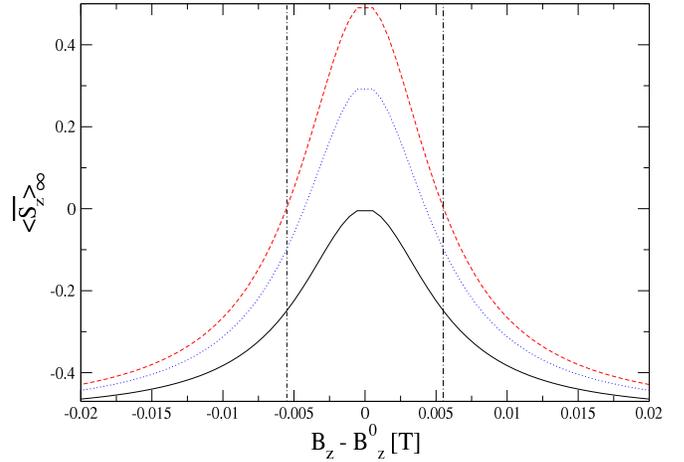}

\caption{\label{cap:SzInfResonance}$\overline{\left\langle S_{z}\right\rangle }_{\infty}$
evaluated within Born approximation near the resonance, from Eq. \prettyref{eq:SzStatLimit}
where $B_{z}^{0}=-\frac{pA}{2g^{*}\mu_{B}}$. We have used the value
of $A$ for GaAs, $g^{*}=-0.44$, $N=10^{5}$, and $I=\frac{1}{2}$.
$\left\langle S_{z}\right\rangle _{0}=-\frac{1}{2}$ for all three
curves and results are given for $p=0$ (solid line), $p=\frac{6}{10}$
(dotted line), and $p=1$ (dashed line). The vertical dash-dotted
lines indicate the magnetic fields where the relevant smallness parameter
is unity: $\left|\delta\right|=1$.}
\end{figure}
Before applying the continuum limit, the stationary limit for $\left\langle S_{z}\right\rangle _{t}$
is\begin{equation}
\overline{\left\langle S_{z}\right\rangle }_{\infty}=\frac{\left\langle S_{z}\right\rangle _{0}+\frac{1}{4}\left(c_{-}-c_{+}\right)N_{\mathrm{tot}}}{1+\frac{c_{+}+c_{-}}{2}N_{\ \mathrm{tot}}}.\end{equation}
 After applying the continuum limit, $N_{\mathrm{tot}}\to\infty$,
we obtain\begin{equation}
\overline{\left\langle S_{z}\right\rangle }_{\infty}=\frac{1}{2}\frac{c_{-}-c_{+}}{c_{-}+c_{+}}=\frac{pI}{c_{-}+c_{+}}.\label{eq:SzInfinityResonance}\end{equation}
 For $I=\frac{1}{2}$, $\overline{\left\langle S_{z}\right\rangle }_{\infty}=\frac{p}{2}$,
which appears to be an intuitive result. However, evaluating the remaining
pole contributions at the resonance, we find, for a two-dimensional
quantum dot, \begin{equation}
2\mathrm{Re}\left[P_{3}(t)\right]=\left[\left\langle S_{z}\right\rangle _{0}-\frac{2pI}{c_{-}+c_{+}}\right]\cos\left(\Omega_{0}t\right)+O\left(\frac{1}{N}\right),\label{eq:SzPolesResonance}\end{equation}
 where\begin{equation}
\Omega_{0}=\sqrt{\frac{N}{2}\left(c_{+}+c_{-}\right)}.\label{eq:Omega0Resonance}\end{equation}
 The results in \prettyref{eq:SzInfinityResonance} and \prettyref{eq:SzPolesResonance}
do not reproduce the exact solution in the limit $p=1$, $I=\frac{1}{2}$,
and do not recover the correct $t=0$ value of $\left\langle S_{z}\right\rangle _{t}$.
The Born approximation for $\left\langle S_{z}\right\rangle _{t}$,
as it has been defined here, breaks down in the strongly non-perturbative
limit, although the transverse components are better behaved.

On resonance, the poles at $s_{1}$ and $s_{3}$ are equidistant from
the origin, and the major contributions to $\left\langle S_{+}\right\rangle _{t}$
come from these two poles: $\left\langle S_{+}\right\rangle _{t}\approx P_{1}(t)+P_{3}(t)$.
Evaluating the residues at these poles,\begin{equation}
\left\langle S_{+}\right\rangle _{t}=\left\langle S_{+}\right\rangle _{0}\left(1-O\left(\frac{1}{N}\right)\right)\cos\left(\Omega_{0}t\right),\end{equation}
 which suggests that a fraction $O\left(\frac{1}{N}\right)$ of the
spin undergoes decay, and the rest precesses at a frequency $\Omega_{0}$.
When $I=\frac{1}{2}$, and in proper energy units we have $\Omega_{0}=\frac{A}{\sqrt{8N}}$
from Eq. \prettyref{eq:Omega0Resonance}. While it does not violate
positivity, as in the case of $\left\langle S_{z}\right\rangle _{t}$,
this expression should not be taken seriously in general, since this
result has been obtained well outside of the perturbative regime.
The above does, however, recover the exact solution in the limit $p=1$.
We show the stationary limit of $\left\langle S_{z}\right\rangle _{t}$
in Fig. \prettyref{cap:SzInfResonance}, using typical values for
an electron confined to a GaAs quantum dot.

\subsection{Fourth-order corrections}

The fourth order expansion of the self-energy for $\left\langle S_{z}\right\rangle _{t}$
is given in Appendix \ref{sec:Self-energy-expansion}. The discrete
expression for the numerator term $N_{z}^{(4)}(s)$ contains second
order poles (secular terms). The fourth-order expression for $S_{z}(s)$
inherits these second order poles (see Eq. \prettyref{eq:SzLaplaceTransform}).
When the Laplace transform is inverted, this will result in pole contributions
that grow linearly in time. However, when the continuum limit is performed,
which is strictly valid for times shorter than $t\approx\sqrt{N}$
(see Appendix \ref{sec:Continuum-limit}), all poles in $N_{z}^{(4)}(s)$
are replaced by branch cuts. The integrals around the branch cuts
can then be performed to obtain a solution for $\left\langle S_{z}\right\rangle _{t}$,
valid for times $t\lesssim\sqrt{N}$.

All relevant non-analytic features (branch points and poles) of $S_{z}(s)$
occur in two regions of the complex plane: about the origin $s\approx0$,
and at high frequencies, around $s\approx\pm i\omega_{n}$. Inserting
an initial nuclear state $\ket{n}$ for a large uniform system (see
Appendix \ref{sec:InitStateUnifPolarization}), expanding the fourth-order
self-energy to leading order in $\frac{1}{\omega_{n}}$ about the
points $s=0$ and $s=-i\omega_{n}$, performing the continuum limit,
and evaluating the integrals over coupling constants, we obtain (where
the overbar and {}``conj.'' indicate complex conjugate for $s$
real):\begin{widetext}

\begin{eqnarray}
N_{z}^{(4)}(s-i\omega_{n}) & \simeq & -\frac{\Delta^{2}}{2}\left\{ c_{+}c_{-}\left[L_{1}(s)+L_{2}(s)-L_{3}(s)-\mathrm{conj}.\right]+c_{+}^{2}L_{1}(s)-c_{-}^{2}\bar{L}_{1}(s)\right\} \\
\Sigma_{zz}^{(4)}(s-i\omega_{n}) & \simeq & -N\Delta\left\{ c_{+}c_{-}\left[L_{1}(s)+L_{2}(s)-L_{3}(s)+\mathrm{conj}.\right]+c_{+}^{2}L_{1}(s)+c_{-}^{2}\bar{L}_{1}(s)\right\} \\
N_{z}^{(4)}(s) & \simeq & \frac{\delta^{2}}{2}\left(c_{+}^{2}-c_{-}^{2}\right)\left(\frac{3}{4}+s^{2}L_{4}(s)\right)\\
\Sigma_{zz}^{(4)}(s) & \simeq & is\delta^{2}\left[3\left(pI\right)^{2}+\left(c_{+}^{2}+c_{-}^{2}+14c_{+}c_{-}\right)s^{2}L_{4}(s)\right]\end{eqnarray}
 with coupling constant integrals $L_{i}(s)$ given by

\begin{eqnarray}
L_{1}(s) & = & \frac{i}{2(s+i)}-\frac{1}{2}\left[\log(s+i)-\log(s)\right]\\
L_{3}(s) & = & \left[s\log(s+i)-s\log(s)-i\right]^{2}\\
L_{4}(s) & = & \frac{1}{6}-\frac{1}{6s}\left[s^{3}+3s+2i\right]\left[\log(s+i)-\log(s)\right]-\frac{1}{6s}\left[s^{3}+3s-2i\right]\left[\log(s-i)-\log(s)\right]\end{eqnarray}
 and\begin{multline}
L_{2}(s)=\log(s+i)-\log(s)-i\left[(s+i)\log(s+i)-(s+2i)\log(s+2i)+s\log(s)-(s-i)\log(s-i)\right]\\
+is\int_{s}^{s+i}du\frac{\log(2u-s-i)-\log(2u-s)}{u}.\end{multline}

Noting that $\lim_{s\to0}s^{2}L_{4}(s)=0$, we find the corrections
to the stationary limit for $\left\langle S_{z}\right\rangle _{t}$.
At fourth order in the flip-flop terms, this gives\begin{equation}
\overline{\left\langle S_{z}\right\rangle }_{\infty}=\frac{\left\langle S_{z}\right\rangle _{0}+pI\delta+\frac{3}{8}\left(c_{+}^{2}-c_{-}^{2}\right)\delta^{2}+O\left(\frac{N}{\omega_{n}^{4}}\right)}{1+(c_{+}+c_{-})\delta-3(pI)^{2}\delta^{2}+O\left(\frac{N}{\omega_{n}^{4}}\right)}.\label{eq:SzInfFourthOrder}\end{equation}

\end{widetext}

\begin{table*}
\begin{tabular}{|c|c|c|c|c||c|c|}
\hline 
&
$\Sigma_{S}\simeq\Sigma_{S}^{(2)}$&
$\Sigma_{S}\simeq\Sigma_{S}^{(2)}$&
$\Sigma_{S}\simeq\Sigma_{S}^{(2)}$&
$\Sigma_{S}\simeq\Sigma_{S}^{(2)}+\Sigma_{S}^{(4)}$&
$\Sigma_{S}\simeq\Sigma_{S}^{(2)}$&
$\Sigma_{S}\simeq\Sigma_{S}^{(2)}$\tabularnewline
\hline 
&
$b^{\prime}=0,$&
$b^{\prime}\ne0,$&
$b^{\prime}\ne0,$&
$p\ne1,$&
$\left|\Delta\right|\ll1,$&
$\left|\Delta\right|\ll1,$\tabularnewline
&
$\frac{d}{m}=1,$&
$\frac{d}{m}=1,$&
$\frac{d}{m}=1,$&
$\frac{d}{m}=1,$&
$\frac{d}{m}<2,$&
$\frac{d}{m}\ge2,$\tabularnewline
\multicolumn{1}{|c|}{}&
$t\gg1$&
$\Gamma_{2}^{-1}\gtrsim t\gg1$&
$t\gg e^{\left|b^{\prime}\right|/N}\gg\Gamma_{2}^{-1}$&
$t\gg e^{\left|b^{\prime}\right|/N},\, b^{\prime}\gg2pIN$&
$t\gg1$&
$t\gg1$\tabularnewline
\hline
\hline 
$R_{X}(t)\propto$&
$1/\ln t$&
$e^{i\omega_{2}t}e^{-\Gamma_{2}t}$&
$1/t\ln^{2}t$&
$1/t\ln^{3}t$,~$X=z$&
$(1/t)^{\frac{d}{m}}e^{\pm it}$&
$\ln^{\nu}t/t^{2},\,\,\,\,\,\nu=\frac{d}{m}-1$\tabularnewline
\hline
\end{tabular}

\caption{\label{cap:ResultsTable}Results for the decaying fraction of the
spin $\left(\left|R_{X}(t)\right|<O(\delta)\,\forall\, t\right)$
in various parameter regimes. Results are given for both remainder
terms $R_{X}(t),\, X=z,+$, within the Born approximation for the
self-energy $\Sigma_{S}\simeq\Sigma_{S}^{(2)}$ and for $R_{z}(t)$
at fourth order in the nuclear spin--electron spin flip-flop terms
$\Sigma_{S}\simeq\Sigma_{S}^{(2)}+\Sigma_{S}^{(4)}$ when $p\ne1$.
The first three columns are exact in the limit of full polarization
$(p=1)$ of the nuclear spin system, but still may describe the correct
electron spin dynamics in the weakly perturbative regime, $\left|\Delta\right|\lesssim1$.
The last two columns give the correct electron spin dynamics in the
strongly perturbative regime, $\left|\Delta\right|\ll1$. }
\end{table*}

The fourth-order corrections to the self-energy at high frequency
($s\approx-i\omega_{n}$) are suppressed relative to the Born approximation
by an additional factor of the smallness parameter $\Delta$, as expected
from the analysis given in Appendix \ref{sec:Self-energy-expansion}.
However, the low-frequency ($s\approx0$) part of the fourth-order
self-energy is suppressed by the much smaller parameter $\delta$.
This allows us to determine the stationary limit of $\left\langle S_{z}\right\rangle _{t}$
with confidence even when the magnetic field is small or zero, provided
the polarization is sufficiently large. When $b^{\prime}=0$ and $I=\frac{1}{2}$,
we have $\delta=\frac{1}{p^{2}N}$, so the stationary limit can be
determined whenever $p\gg\frac{1}{\sqrt{N}}$. 

It is relatively straightforward to find the time-dependence as $t\to\infty$
for the $S_{z}$ branch cut integrals at fourth order. Neglecting
contributions from the branch cuts near $s\simeq0$, which are suppressed
by the factor $\delta^{2}$, and when $p<1$ so that the coefficient
$c_{+}\ne0$ (c.f. Eq. \prettyref{eq:CPlusMinusDefinition}), we find
the major contributions at long times come from the branch points
at $s=\pm i$, where $L_{2}(s)\propto\log^{2}(s+i)$. For any magnetic
field, we find:\begin{equation}
R_{z}\left(t\to\infty\right)\propto\frac{1}{t\ln^{3}t}.\end{equation}
For $b^{\prime}\gg2pIN$, this time-dependence will be dominant when
$t\gg\exp\left(\frac{\left|b^{\prime}\right|}{N}\right)$. Thus, we
find that the fourth-order result has a faster long-time decay than
the Born approximation, and that the associated asymptotics are valid
at the same times as the Born approximation asymptotics (see Eq. \prettyref{eq:BranchCutsNonzeroMagField}).
Thus, higher-order corrections may change the character of the long-time
decay in the weakly perturbative regime, where they are not negligible.
In contrast, in the strongly perturbative regime $\left|\Delta\right|\ll1$,
the fourth- and higher-order terms are negligible, so the Born approximation
dominates for all times $t<\exp\left(\left|b^{\prime}\right|/N\right)$.

\section{Conclusions\label{sec:Conclusions}}

We have given a complete analytical description for the dynamics of
an electron spin interacting with a nuclear spin environment via the
Fermi contact hyperfine interaction. In a large magnetic field, our
calculation applies to a nuclear spin system of arbitrary polarization
$p$ and arbitrary spin $I$, prepared in an eigenstate of the total
$z$-component of the (quantum) nuclear Overhauser field. In the limit
of full polarization $p=1$ and nuclear spin $I=\frac{1}{2}$, the
Born approximation applied to the self-energy recovers the exact dynamics
for $\left\langle S_{+}\right\rangle _{t}$ and $\left\langle S_{z}\right\rangle _{t}$,
with all non-perturbative effects. We have shown explicitly that the
dynamical behavior we calculate in Born approximation is purely non-Markovian,
and can be obtained in the limit of high magnetic fields directly
from the remainder term to a Born-Markov approximation. By performing
our expansion on the self-energy superoperator, we have re-summed
secular divergences that are present in standard perturbation theory
at lowest (second) order for the transverse components $\left\langle S_{+}\right\rangle _{t}$
and at fourth and higher order for the longitudinal spin $\left\langle S_{z}\right\rangle _{t}$.
For low magnetic fields $b^{\prime}\lesssim N$ , but still within
the perturbative regime ($\left|\Delta\right|<1$), the Born approximation
for the electron spin shows rich dynamics including non-exponential
(inverse logarithm) decay, exponential decay, and undamped oscillations.
For high magnetic fields $b^{\prime}\gg N$, and for $\frac{d}{m}<2$,
the electron spin shows a power-law decay ($\sim\left(\frac{1}{t}\right)^{\frac{d}{m}}$
in d-dimensions for an isotropic envelope wave function of the form
$\psi(r)\propto\exp\left[-\frac{1}{2}\left(\frac{r}{l_{0}}\right)^{m}\right]$)
to its stationary value with a time scale $\tau_{\mathrm{hf}}\approx\frac{2N\hbar}{A}$,
in agreement with the exact solution for a fully polarized nuclear
spin system.\cite{khaetskii:2002a,khaetskii:2003a} Above a critical
ratio, $\frac{d}{m}\ge2$, the spin decay asymptotics undergo an abrupt
change, signaled by a disappearance of slow oscillations in the decay
envelope. We have summarized these results in Table \prettyref{cap:ResultsTable}.
We have also suggested a method that could be used to probe the non-Markovian
electron spin dynamics directly, using a standard spin-echo technique.
We emphasize that the electron spin only decays by some small fraction
of its initial value, of order $\delta$ (see Tables \prettyref{cap:Symbols},
\prettyref{cap:SymbolsNumericalValues}), and the decay is generically
non-exponential at long times (see Table \prettyref{cap:ResultsTable}).
The results of this work may therefore be of central importance to
the development of future quantum error correction schemes, which
typically assume an exponential decay to zero. The fact that the stationary
value of the spin depends on the initial value implies that this system
is non-ergodic. Based on this observation, we postulate a general
principle, that non-ergodic quantum systems can preserve phase-coherence
to a higher degree than systems with ergodic behavior. It would be
interesting to explore this connection further.

\begin{acknowledgments}
We thank B. L. Altshuler, O. Chalaev, H.-A. Engel, S. Erlingsson,
H. Gassmann, V. Golovach, A. V. Khaetskii, F. Meier, D. S. Saraga,
J. Schliemann, and E. A. Yuzbashyan for useful discussions. We acknowledge
financial support from the Swiss NSF, the NCCR nanoscience, EU RTN
Spintronics, DARPA, ARO, and ONR. WAC acknowledges funding from NSERC
of Canada.
\end{acknowledgments}
\appendix

\section{Self-energy expansion\label{sec:Self-energy-expansion}}

To expand the self-energy superoperator $\Sigma_{S}$ in powers of
$L_{V}$, we have found it convenient to work in terms of a superoperator
matrix representation. Here we give a brief description of its use
and apply it to generate the reduced self-energy at second order in
$L_{V}$ for all spin components, and the fourth order for the longitudinal
spin. 

Any operator $\mathcal{O}$ that acts on both the electron spin and
nuclear spin Hilbert spaces can be written in terms of the $2\times2$
identity, $\sigma_{0}$, and the Pauli matrices $\sigma_{i},\, i=(x,y,z)$:\begin{equation}
\mathcal{O}=\sum_{i=(0,x,y,z)}c_{i}\sigma_{i}\end{equation}
 where the coefficients $c_{i}$ are operators that act only on the
nuclear spin space. Equivalently, $\mathcal{O}$ can be written in
terms of the operators $\rho_{\uparrow/\downarrow}=\frac{1}{2}\left(\sigma_{0}\pm\sigma_{z}\right)$,
$S_{\pm}=\frac{1}{2}\left(\sigma_{x}\pm i\sigma_{y}\right)$, i.e.:\begin{equation}
\mathcal{O}=k_{\uparrow}\rho_{\uparrow}+k_{\downarrow}\rho_{\downarrow}+k_{+}S_{-}+k_{-}S_{+}\label{eq:OperatorOinRhoBasis}\end{equation}
 with operators $k_{j}$ that act on the nuclear spin space. We have
labeled the coefficients $k_{j}$ in this way so that when $\mathcal{O}=\rho_{S}$
is the electron spin density operator, $k_{\pm}=\left\langle S_{\pm}\right\rangle $.
A superoperator $\mathcal{S}$ acting on $\mathcal{O}$ maps it to
the operator $\mathcal{O}^{\prime}$ with new coefficients:\begin{equation}
\mathcal{SO}=\mathcal{O}^{\prime}=k_{\uparrow}^{\prime}\rho_{\uparrow}+k_{\downarrow}^{\prime}\rho_{\downarrow}+k_{+}^{\prime}S_{-}+k_{-}^{\prime}S_{+}.\label{eq:SOinRhoBasis}\end{equation}
 This allows us to write $\mathcal{O}$ as a vector and $\mathcal{S}$
as a $4\times4$ matrix, the elements of which are superoperators
that act on the nuclear spin space, and are determined by \prettyref{eq:OperatorOinRhoBasis}
and \prettyref{eq:SOinRhoBasis}:\begin{eqnarray}
\vec{\mathcal{O}} & = & \left(k_{\uparrow},k_{\downarrow},k_{+},k_{-}\right)^{T}\\
\vec{\mathcal{O}}^{\prime} & = & \left(k_{\uparrow}^{\prime},k_{\downarrow}^{\prime},k_{+}^{\prime},k_{-}^{\prime}\right)^{T},\end{eqnarray}
 $\vec{\mathcal{O}}^{\prime}=\left[\mathcal{S}\right]\vec{\mathcal{O}}$
and $k_{\alpha}^{\prime}=\sum_{\beta}\mathcal{S}_{\alpha\beta}k_{\beta}$,
where $\alpha,\,\beta=\uparrow,\downarrow,+,-$. 

Laplace transforming the reduced self-energy given in \prettyref{eq:ReducedNZGME}
yields\begin{eqnarray}
\Sigma_{S}(s) & = & -i\mathrm{Tr}_{I}L\frac{1}{s+iQL}L_{V}\rho_{I}(0),\end{eqnarray}
 which is expanded in powers of $L_{V}$

\begin{equation}
\frac{1}{s+iQL}=\sum_{k=0}^{\infty}\frac{1}{s+iQL_{0}}\left(-iQL_{V}\frac{1}{s+iQL_{0}}\right)^{k}.\end{equation}
To obtain these higher order terms in the self-energy, we form products
of the free propagator $\frac{1}{s+iQL_{0}}$ and the perturbation
$QL_{V}$. The free propagator is diagonal in the basis of $\left\{ \rho_{\uparrow/\downarrow},\, S_{\pm}\right\} $,
and is given in terms of \textbf{$2\times2$} blocks by\begin{eqnarray}
\left[\frac{1}{s+iQL_{0}}\right] & = & \left(\begin{array}{cc}
G_{I}^{0}(s) & 0\\
0 & G_{I}^{0\prime}(s)\end{array}\right),\end{eqnarray}
 where \begin{eqnarray}
G_{I}^{0}(s) & = & \left(\begin{array}{cc}
G_{\uparrow}^{0}(s) & 0\\
0 & G_{\downarrow}^{0}(s)\end{array}\right),\\
G_{I}^{0\prime}(s) & = & \left(\begin{array}{cc}
G_{+}^{0}(s) & 0\\
0 & G_{-}^{0}(s)\end{array}\right).\end{eqnarray}
 In the above,\begin{eqnarray}
G_{\uparrow/\downarrow}^{0}(s) & = & \frac{1}{s\pm i\frac{Q}{2}L_{\omega}^{-}},\\
G_{\pm}^{0}(s) & = & \frac{1}{s\mp i\frac{Q}{2}L_{\omega}^{+}},\end{eqnarray}
 where we define the new (nuclear spin) Liouvillians by their action
on an arbitrary operator $\mathcal{O}$: $L_{\omega}^{\pm}\mathcal{O}=\commute{\omega}{\mathcal{O}}_{\pm},\,\omega=b^{\prime}+h_{z}$.
The perturbation term contains only off-diagonal elements when written
in terms of $2\times2$ blocks:\begin{eqnarray}
\left[QL_{V}\right] & = & \left(\begin{array}{cc}
0 & V_{I}\\
V_{I}^{\prime} & 0\end{array}\right),\end{eqnarray}
 where we find

\begin{equation}
V_{I}=\frac{Q}{2}\left(\begin{array}{cc}
h_{-}^{L} & -h_{+}^{R}\\
-h_{-}^{R} & h_{+}^{L}\end{array}\right),\,\,\,\,\, V_{I}^{\prime}=\frac{Q}{2}\left(\begin{array}{cc}
h_{+}^{L} & -h_{+}^{R}\\
-h_{-}^{R} & h_{-}^{L}\end{array}\right).\end{equation}
In the above expression, we have introduced superoperators for right
and left multiplication: \begin{eqnarray}
\mathcal{O}^{R}\mathcal{A} & = & \mathcal{AO}\\
\mathcal{O}^{L}\mathcal{A} & = & \mathcal{O}\mathcal{A}.\end{eqnarray}

Only even powers of $L_{V}$ can contribute to the final trace over
the nuclear system, so we consider a general term in the expansion
of the self-energy\begin{eqnarray}
\left(\left[QL_{V}\right]\left[\frac{1}{s+iQL_{0}}\right]\right)^{2k} & = & \left(\begin{array}{cc}
\Sigma_{k} & 0\\
0 & \Sigma_{k}^{\prime}\end{array}\right),\end{eqnarray}

\begin{equation}
\Sigma_{k}=\left(V_{I}G_{I}^{0\prime}V_{I}^{\prime}G_{I}^{0}\right)^{k},\,\,\,\,\,\Sigma_{k}^{\prime}=\left(V_{I}^{\prime}G_{I}^{0}V_{I}G_{I}^{0\prime}\right)^{k}.\end{equation}
 By inspection of the form of $V_{I}$, $V_{I}^{\prime}$, we find
that the $2\times2$ matrix $\mathrm{Tr}_{I}\Sigma_{k}^{\prime}\rho_{I}(0)$
is diagonal when $\rho_{I}(0)=\ket{n}\bra{n}$, and $\ket{n}$ is
an eigenstate of $h_{z}$ (as in Eq. \prettyref{eq:SpecificInitialState}),
since the off-diagonal components always contain terms proportional
to $h_{+}^{2}$ or $h_{-}^{2}$. Thus, to all orders in the perturbation
$L_{V}$, the reduced self-energy takes the form\begin{equation}
\Sigma_{S}(s)=\left(\begin{array}{cccc}
\Sigma_{\uparrow\uparrow}(s) & \Sigma_{\uparrow\downarrow}(s) & 0 & 0\\
\Sigma_{\downarrow\uparrow}(s) & \Sigma_{\downarrow\downarrow}(s) & 0 & 0\\
0 & 0 & \Sigma_{++}(s) & 0\\
0 & 0 & 0 & \Sigma_{--}(s)\end{array}\right).\end{equation}
 The number of matrix elements left to calculate can be further reduced
with the relationships $\Sigma_{\uparrow\uparrow}(s)=-\Sigma_{\downarrow\uparrow}(s)$,
$\Sigma_{\uparrow\downarrow}(s)=-\Sigma_{\downarrow\downarrow}(s)$,
which follow directly from the condition $\mathrm{Tr}\dot{\rho}_{S}=0\Rightarrow\dot{\rho}_{\uparrow}(t)=-\dot{\rho}_{\downarrow}(t)$
and the GME $\dot{\rho}_{\alpha}=-i\sum_{\beta=\uparrow,\downarrow}\int_{0}^{t}dt^{\prime}\Sigma_{\alpha\beta}(t-t^{\prime})\rho_{\beta}(t^{\prime}),\,\alpha=\uparrow,\downarrow$.
By direct calculation we find

\begin{widetext}

\begin{eqnarray}
\Sigma_{\uparrow\uparrow}^{(2)}(s) & = & -\frac{i}{4}\sum_{k}\left[h_{-}\right]_{nk}\left[h_{+}\right]_{kn}\left(\frac{1}{s-i\bar{\omega}_{nk}}+\frac{1}{s+i\bar{\omega}_{nk}}\right)\label{eq:SigUpUpBornAnyn}\\
\Sigma_{\uparrow\downarrow}^{(2)}(s) & = & \frac{i}{4}\sum_{k}\left[h_{+}\right]_{nk}\left[h_{-}\right]_{kn}\left(\frac{1}{s-i\bar{\omega}_{nk}}+\frac{1}{s+i\bar{\omega}_{nk}}\right)\label{eq:SigUpDownBornAnyn}\\
\Sigma_{++}^{(2)}(s) & = & -\frac{i}{4}\sum_{k}\left(\left[h_{+}\right]_{nk}\left[h_{-}\right]_{kn}+\left[h_{-}\right]_{nk}\left[h_{+}\right]_{kn}\right)\frac{1}{s-i\delta\omega_{nk}}.\label{eq:SigPlusPlusBornAnyn}\end{eqnarray}
 In the above, $\bar{\omega}_{nk}=\frac{1}{2}\left(\omega_{n}+\omega_{k}\right)$,
$\delta\omega_{nk}=\frac{1}{2}\left(\omega_{n}-\omega_{k}\right)$,
and $\omega_{j}=b^{\prime}+\left[h_{z}\right]_{jj}$. At fourth order, 

\begin{multline}
\Sigma_{\uparrow\uparrow}^{(4)}(s)=\frac{i}{16}\left\{ \sum_{k_{1}k_{2}k_{3}}\left[h_{-}\right]_{nk_{3}}\left[h_{+}\right]_{k_{3}k_{2}}\left[h_{-}\right]_{k_{2}k_{1}}\left[h_{+}\right]_{k_{1}n}\left((1-\delta_{nk_{2}})\sigma_{4A}^{k_{1}k_{2}k_{3}}(s)+\sigma_{4B}^{k_{1}k_{2}k_{3}}(s)\right)\right.\\
\left.-\frac{1}{s}\sum_{k_{1}k_{2}}\left[h_{-}\right]_{nk_{2}}\left[h_{+}\right]_{k_{2}n}\left[h_{+}\right]_{nk_{1}}\left[h_{-}\right]_{k_{1}n}\sigma_{4C}^{k_{1}k_{2}}(s)\right\} ,\end{multline}

\begin{multline}
\Sigma_{\uparrow\downarrow}^{(4)}(s)=-\frac{i}{16}\left\{ \sum_{k_{1}k_{2}k_{3}}\left[h_{+}\right]_{nk_{3}}\left[h_{-}\right]_{k_{3}k_{2}}\left[h_{+}\right]_{k_{2}k_{1}}\left[h_{-}\right]_{k_{1}n}\left((1-\delta_{nk_{2}})\bar{\sigma}_{4A}^{k_{1}k_{2}k_{3}}(s)+\bar{\sigma}_{4B}^{k_{1}k_{2}k_{3}}(s)\right)\right.\\
\left.-\frac{1}{s}\sum_{k_{1}k_{2}}\left[h_{+}\right]_{nk_{2}}\left[h_{-}\right]_{k_{2}n}\left[h_{-}\right]_{nk_{1}}\left[h_{+}\right]_{k_{1}n}\sigma_{4C}^{k_{1}k_{2}}(s)\right\} ,\end{multline}
where the overbar indicates complex conjugation for $s$ real and
\begin{multline}
\sigma_{4A}^{k_{1}k_{2}k_{3}}(s)=\frac{1}{s-i\delta\omega_{nk_{2}}}\frac{1}{s-i\bar{\omega}_{nk_{1}}}\left(\frac{1}{s-i\bar{\omega}_{nk_{3}}}+\frac{1}{s+i\bar{\omega}_{k_{2}k_{3}}}\right)\\
+\frac{1}{s+i\delta\omega_{nk_{2}}}\frac{1}{s+i\bar{\omega}_{nk_{3}}}\left(\frac{1}{s+i\bar{\omega}_{nk_{1}}}+\frac{1}{s-i\bar{\omega}_{k_{1}k_{2}}}\right),\end{multline}
\begin{eqnarray}
\sigma_{4B}^{k_{1}k_{2}k_{3}}(s) & = & \frac{1}{s-i\delta\omega_{k_{1}k_{3}}}\left(\frac{1}{s-i\bar{\omega}_{k_{1}k_{2}}}+\frac{1}{s+i\bar{\omega}_{k_{2}k_{3}}}\right)\left(\frac{1}{s+i\bar{\omega}_{nk_{3}}}+\frac{1}{s-i\bar{\omega}_{k_{1}n}}\right),\\
\sigma_{4C}^{k_{1}k_{2}}(s) & = & \frac{4s^{2}}{\left(s^{2}+\bar{\omega}_{nk_{1}}^{2}\right)\left(s^{2}+\bar{\omega}_{nk_{2}}^{2}\right)}.\end{eqnarray}

\end{widetext} 

Every two powers of the perturbation $L_{V}$ are associated with
an additional sum over $\approx N$ nuclear spin sites, since every
spin flip up must be paired with a flop down. Non-analyticities (poles)
of the self-energy occur in two regions of the complex plane: at high
frequencies, near $s\approx\pm i\omega_{n}$, and at low frequency,
around $s\approx0$. Expanding near either of these two points gives
an extra factor $\frac{1}{\omega_{n}}$ for every two orders of $QL_{V}\frac{1}{s-iQL_{0}}$.
The self-energy at $\left(2k\right)^{\mathrm{th}}$ order is then
suppressed at least by the factor $\Delta^{k}$, where $\Delta=\frac{N}{\omega_{n}}$:
\begin{eqnarray}
\Sigma_{S}^{(2k)}(s) & \propto & \Delta^{k},\\
\Sigma_{S}^{(2k)}(s-i\omega_{n}) & \propto & \Delta^{k}.\end{eqnarray}
 Thus, in general, for the perturbation series to be well-controlled,
we require $\left|\Delta\right|\ll1$.

\section{Coefficients $c_{\pm}$\label{sec:InitStateUnifPolarization}}

We are interested in evaluating the expressions given in Eqs. \prettyref{eq:SigUpUpBornAnyn},
\prettyref{eq:SigUpDownBornAnyn}, and \prettyref{eq:SigPlusPlusBornAnyn}.
To do this, we investigate objects of the form\begin{equation}
\sum_{k}\left[h_{\pm}\right]_{nk}\left[h_{\mp}\right]_{kn}f_{\mp}(k),\label{eq:HPlusMinusProduct}\end{equation}
 where $f_{\mp}(k)$ is a function of the state index $k$. Inserting
$\ket{n}$, as given in Eq. \prettyref{eq:SpecificInitialState},
into \prettyref{eq:HPlusMinusProduct}, we find \begin{equation}
\sum_{k}\left[h_{\pm}\right]_{nk}\left[h_{\mp}\right]_{kn}f_{\mp}(k)=\sum_{k}A_{k}^{2}c_{\mp}^{k}f_{\mp}(k),\end{equation}
 where the state index $k$ now labels sites at which a nuclear spin
has been raised or lowered, and with the help of the matrix elements:
$\bra{I,m\pm1}I^{\pm}\ket{I,m}=\sqrt{(I\mp m)(I\pm m+1)}$, we have\begin{equation}
c_{\pm}^{k}=\sum_{j=1}^{g_{n}}\left|\alpha_{j}\right|^{2}\left[I(I+1)-m_{k}^{j}(m_{k}^{j}\pm1)\right].\label{eq:CPlusMinusk}\end{equation}
 We assume the initial nuclear system is uniform, so that $c_{\pm}^{k}$
is independent of the site index $k$, and for a large number of degenerate
states $g_{n}\gg1$ that contribute to $\ket{n}$, we replace the
sum over weighting factors $\left|\alpha_{j}\right|^{2}$ by an appropriate
probability distribution. This gives $c_{\pm}^{k}=c_{\pm}$ with $c_{\pm}$
defined in Eq. \prettyref{eq:CPlusMinusDefinition} of the main text.

\section{Continuum limit\label{sec:Continuum-limit}}

Here, we find a rigorous bound on corrections to the memory kernels,
after we have changed sums to integrals. We consider the real-time
version of the functions $I_{\pm}(s)$, given in \prettyref{eq:IPlusMinusLaplaceTrans},
with coupling constants for a Gaussian wave function in two dimensions
($m=d=2$ in Eq. \prettyref{eq:AkDefinition}):\begin{equation}
I_{\pm}(t)=\frac{1}{4N}\sum_{k}A_{k}^{2}e^{\pm iA_{k}t/2},\, A_{k}=2e^{-k/N}.\end{equation}
 The Euler-MacLauren formula gives an upper bound to the corrections
involved in the transformation of sums to integrals for a summand
that is a smooth monotonic function of its argument. For times $t\gg1$,
the summand of $I_{\pm}(t)$ is not monotonic on the interval $k=1,\ldots,N$,
where it has appreciable weight. We divide the sum into $t$ subintervals
of width $\Delta k\approx\frac{N}{t}$. The summand is then monotonic
over each of the $t$ subintervals, and the Euler-MacLauren formula
gives a remainder $R\le\frac{2}{N}$ when the sum over each subinterval
is changed to an integral. Adding the errors incurred for each subinterval,
we find (for $t\gg1$):\begin{equation}
I_{\pm}(t)=\left[\left(e^{\pm it}-1\right)\frac{1}{t^{2}}\mp\frac{ie^{\pm it}}{t}\right]+R(t).\end{equation}
 The remainder term $\left|R(t)\right|\le\frac{2t}{N}$, so the corrections
can become comparable to the amplitude of the integral itself when
$t\approx\sqrt{N/2}$. This represents a strict lower bound to the
time scale where the continuum limit is valid for $m=d=2$.

\section{Perturbation theory\label{sec:Perturbation-theory}}

In this Appendix we apply standard perturbation theory to the problem
of finding the electron spin dynamics. We do this to illustrate the
connection between our perturbative expansion of the self-energy and
the standard one, and to demonstrate the need for a non-perturbative
approach.

We choose the initial state 

\begin{equation}
\ket{i}=\left(\sqrt{\rho_{\uparrow}(0)}\ket{\uparrow}+e^{i\phi}\sqrt{1-\rho_{\uparrow}(0)}\ket{\downarrow}\right)\otimes\ket{n},\end{equation}
 where $\ket{n}$ is an eigenstate of $h_{z}$ and $\left\langle S_{z}\right\rangle _{0}=\frac{1}{2}\left(\rho_{\uparrow}(0)-\rho_{\downarrow}(0)\right),\,\left\langle S_{+}\right\rangle _{0}=\sqrt{\rho_{\uparrow}(0)\left(1-\rho_{\downarrow}(0)\right)}e^{i\phi}$.
We then apply standard interaction picture perturbation theory to
evaluate $\left\langle S_{X}\right\rangle _{t},\, X=+,z$. To lowest
nontrivial (second) order in the perturbation $\mathcal{H}_{V}$,
we find\begin{widetext}

\begin{eqnarray}
\left\langle S_{+}\right\rangle _{t} & = & e^{i\omega_{n}t}\left\langle S_{+}\right\rangle _{0}-\frac{\left\langle S_{+}\right\rangle _{0}}{4}\sum_{k}\left(\left[h_{-}\right]_{nk}\left[h_{+}\right]_{kn}g_{k}^{-}(t)+\left[h_{+}\right]_{nk}\left[h_{-}\right]_{kn}g_{k}^{+}(t)\right)\\
g_{k}^{\pm} & = & \frac{te^{i\omega_{n}t}}{i\bar{\omega}_{nk}}-\frac{1}{\bar{\omega}_{nk}^{2}}\left(e^{\pm i\frac{A_{k}}{2}t}-e^{i\omega_{n}t}\right)\\
\left\langle S_{z}\right\rangle _{t} & = & \left\langle S_{z}\right\rangle _{0}+\frac{1}{2}\sum_{k}\left[\left(1-\rho_{\uparrow}(0)\right)\left[h_{+}\right]_{nk}\left[h_{-}\right]_{kn}-\rho_{\uparrow}(0)\left[h_{-}\right]_{nk}\left[h_{+}\right]_{kn}\right]\frac{\left[1-\cos\left(\bar{\omega}_{nk}t\right)\right]}{\bar{\omega}_{nk}^{2}}.\end{eqnarray}
\end{widetext}

The expression for $\left\langle S_{z}\right\rangle _{t}$ has been
given previously,\cite{khaetskii:2002a,khaetskii:2003a} where it
was noted that the perturbative expression for the transverse components
$\left\langle S_{+}\right\rangle _{t}$ contains a term that grows
unbounded in time (as above). Inserting an initial nuclear state $\ket{n}$
with uniform polarization, performing the continuum limit, and expanding
to leading order in $\frac{1}{\omega_{n}}$ gives the final result,
presented in Eqs. \prettyref{eq:PTSplus} and \prettyref{eq:PTSz}.

\section{Branch cut asymptotics\label{sec:Branch-cut-asymptotics}}

\subsection{Long times}

Here we give explicit expressions for the leading-order terms in asymptotic
expansions of the branch cut integrals for long times

\begin{widetext}

\begin{eqnarray}
K_{0}^{z}(t\to\infty) & = & -i\frac{\pi}{(c_{+}+c_{-})N}\left[2\left\langle S_{z}\right\rangle _{0}-\Delta(c_{+}+c_{-})\right]\frac{1}{\ln t}+O\left(\frac{1}{\ln^{2}t}\right),\, b^{\prime}=0,\\
K_{0}^{z}(t\to\infty) & = & \frac{i\pi}{b^{\prime}}\left[\frac{N}{b^{\prime}}(c_{+}+c_{-})\left[2\left\langle S_{z}\right\rangle _{0}-\Delta\right]-2pI\Delta\right]\frac{1}{t^{2}}+O\left(\frac{1}{t^{3}}\right),\, b^{\prime}\ne0,\\
K_{\pm}^{z}(t\to\infty) & = & \mp\frac{\pi e^{\pm it}}{Nc_{\mp}}\left[2\left\langle S_{z}\right\rangle _{0}\mp\Delta\left(\frac{b^{\prime}}{N}\pm(c_{+}+c_{-})\mp c_{\pm}2\ln2\right)\right]\frac{1}{t\ln^{2}t}+O\left(\frac{1}{t\ln^{3}t}\right),\end{eqnarray}

\begin{eqnarray}
K_{0}^{+}(t\to\infty) & = & -i\frac{2\pi}{(c_{+}+c_{-})N}\left\langle S_{+}\right\rangle _{0}\frac{1}{\ln t}+O\left(\frac{1}{\ln^{2}t}\right),\, b^{\prime}=0,\\
K_{0}^{+}(t\to\infty) & = & i2\pi(c_{+}+c_{-})\frac{N}{\left(b^{\prime}\right)^{2}}\left\langle S_{+}\right\rangle _{0}\frac{1}{t^{2}}+O\left(\frac{1}{t^{3}}\right),\, b^{\prime}\ne0,\\
K_{\pm}^{+}(t\to\infty) & = & \mp\frac{2\pi e^{\pm it}}{Nc_{\mp}}\left\langle S_{+}\right\rangle _{0}\frac{1}{t\ln^{2}t}+O\left(\frac{1}{t\ln^{3}t}\right).\end{eqnarray}

\end{widetext}

\subsection{High fields }

For asymptotically large magnetic fields, the $x$-dependence of the
denominator term $D(-x+\gamma_{\alpha}\pm i\eta)$ that appears in
the branch cut integrals (Eq. \prettyref{eq:BCIntegralExpCutoff})
is dominated by the constant contribution $\sim-i\omega_{n}$, except
at very large values of $x$, where it may be that $D(-x+\gamma_{\alpha}\pm i\eta)\approx0$.
We expand the numerator $N_{z}(s_{\alpha}^{X}(x)\pm i\eta)$ and denominator
in $\frac{1}{x}$, retaining terms up to $\mathcal{O}\left(1\right)$
in the numerator and $\mathcal{O}\left(\frac{1}{x}\right)$ in the
denominator. Expanding to leading order in $\frac{1}{x_{0}}\propto\frac{1}{\omega_{n}}$,
except where there is the possibility of a near-singular contribution
$(D\simeq0)$, and assuming $b^{\prime}>0$, we find the branch cut
integrals\begin{eqnarray}
K_{+}^{X}(t) & \simeq & 2\pi i\delta C_{-}^{X}x_{0}\int_{-i}^{-i+\infty}dz\frac{ze^{-zt}}{z-z_{0}}\\
K_{0}^{X}(t) & \simeq & -2\pi i\delta\left(C_{+}^{X}+C_{-}^{X}\right)x_{0}\int_{0}^{\infty}dz\frac{ze^{-zt}}{z-z_{0}}\\
K_{-}^{X}(t) & \simeq & -2\pi i\delta C_{+}^{X}\int_{i}^{i+\infty}dzze^{-zt}\end{eqnarray}
 where $z_{0}$ and $x_{0}$ are defined in Eqs. \prettyref{eq:z0Definition}
and \prettyref{eq:x0Definition}. The coefficients $C_{\pm}^{X}$are
given by Eq. \prettyref{eq:HighFieldCCoeffs}. The sum over all three
branch cut integrals can now be written in terms of two contour integrals

\begin{multline}
\sum_{\alpha=\left(0,+,-\right)}K_{\alpha}^{X}(t)=-2\pi i\delta C_{-}^{X}x_{0}\int_{C^{\prime\prime}}dz\frac{ze^{-zt}}{z-z_{0}}\\
-2\pi i\delta C_{+}^{X}\int_{C^{\prime}}dzze^{-zt}.\end{multline}
$C^{\prime\prime}$ runs clockwise from the origin to $z=\infty$
along the real axis, then returns to $z=-i$, enclosing the pole at
$z=z_{0}$. $C^{\prime}$ runs from $z=i$ to $z=i+\infty$, then
returns along the real axis to $z=0$. These integrals can be evaluated
immediately by closing the contours along the imaginary axis. The
sum of the contributions along the imaginary axis and from the residue
of the pole at $z=z_{0}$ gives the result in Eq. \prettyref{eq:HighFieldBCIntegrals}. 

\bibliographystyle{apsrev}
\bibliography{hfBibliography}

\end{document}